\documentclass{aa} 

\usepackage[T1]{fontenc}
\usepackage[utf8]{inputenc}
\usepackage{microtype}
\usepackage{booktabs}
\usepackage{placeins}

\usepackage{graphicx}
\usepackage{float}

\usepackage{silence}
\usepackage{subcaption}

\usepackage{soul}
\usepackage{txfonts}
\usepackage{hyperref}
\hypersetup{
    colorlinks=true,
    citecolor=blue,
    }

\graphicspath{
	{.}
	{/images/}
	}
\usepackage{xcolor}
\newcommand{\Lagr}{\mathcal{L}}

\makeatletter
\renewcommand*\aa@pageof{, page \thepage{} of \pageref*{LastPage}}
\makeatother

\usepackage{siunitx}
\DeclareSIUnit \parsec {pc}
\DeclareSIUnit \dex {dex}

\let\oldbibliography\thebibliography
\renewcommand{\thebibliography}[1]{%
  \oldbibliography{#1}%
  \setlength{\itemsep}{2pt}%
  \setlength{\baselineskip}{7.5pt}
  \setlength{\lineskiplimit}{-\maxdimen}
}

\begin{document}

   \title{CROCODILE}
   
   \subtitle{Incorporating medium-resolution spectroscopy of\\
   close-in directly imaged exoplanets into atmospheric retrievals\\
   via cross-correlation}

   \author{
        J.~Hayoz\inst{\ref{ethz}}\fnmsep\thanks{\email{jhayoz@ethz.ch}},
        G.~Cugno\inst{\ref{ethz}, \ref{Umich}},
        S.~P.~Quanz\inst{\ref{ethz}},
        P.~Patapis\inst{\ref{ethz}},
        E.~Alei\inst{\ref{ethz}, \ref{nccr}},
        M.~J.~Bonse\inst{\ref{ethz}},
        F.~A.~Dannert\inst{\ref{ethz}, \ref{nccr}},
        E.~O.~Garvin\inst{\ref{ethz}},
        T.~D.~Gebhard\inst{\ref{ethz}, \ref{mpi}},
        B.~S.~Konrad\inst{\ref{ethz}, \ref{nccr}},
        L.F.~Sartori\inst{\ref{ethz}}
   }

    \institute{
        ETH Zurich, Institute for Particle Physics and Astrophysics, Wolfgang-Pauli-Strasse 27, CH-8093 Zurich, Switzerland\label{ethz}
        \and 
        Department of Astronomy, University of Michigan, Ann Arbor, MI 48109, USA\label{Umich}
        \and
        National Center of Competence in Research PlanetS, Switzerland\label{nccr}
        \and
        Max Planck Institute for Intelligent Systems, Max-Planck-Ring 4, 72076 Tübingen, Germany\label{mpi}
    }

    \date{Received 21.12.2022; accepted 28.08.2023}

    \titlerunning{CROCODILE}
    \authorrunning{J. Hayoz et al.} 

    \abstract
   {The investigation of the atmospheres of closely separated, directly imaged gas giant exoplanets is challenging due to the presence of stellar speckles that pollute their spectrum. To remedy this, the analysis of medium- to high-resolution spectroscopic data via cross-correlation with spectral templates (cross-correlation spectroscopy) is emerging as a leading technique.
   }
   {We aim to define a robust Bayesian framework combining, for the first time, three widespread direct-imaging techniques, namely photometry, low-resolution spectroscopy, and medium-resolution cross-correlation spectroscopy in order to derive the atmospheric properties of close-in directly imaged exoplanets. Current atmospheric characterisation frameworks are indeed either not compatible with all three observing techniques or they lack the commitment to efficient sampling strategies that allow high-dimensional forward models.}
   {Our framework \texttt{CROCODILE} (cross-correlation retrievals of directly imaged self-luminous exoplanets) naturally combines the three techniques by adopting adequate likelihood functions. To validate our routine, we simulated observations of gas giants similar to the well-studied $\beta$~Pictoris~b planet and we explored the parameter space of their atmospheres to search for potential biases.
   }
   {We obtain more accurate measurements of atmospheric properties when combining photometry, low- and medium-resolution spectroscopy into atmospheric retrievals than when using the techniques separately as is usually done in the literature. Indeed, the combined fit is, on average, 
   $20\%$ more accurate than fitting only medium-resolution cross-correlation spectroscopy. We find that medium-resolution ($R \approx 4000$) K-band cross-correlation spectroscopy alone is not suitable to constrain the atmospheric properties of our synthetic datasets; 
   however, this problem disappears when simultaneously fitting photometry throughout the Y and M bands and low-resolution ($R \approx 60$) spectroscopy between the Y and H bands. Our thorough testing demonstrates that free chemistry is a suitable forward model to retrieve the atmospheric thermal and chemical properties of cloudless gas giants at chemical equilibrium.}
   {\texttt{CROCODILE} provides a robust statistical framework to interpret medium-resolution spectroscopic data of close-in directly imaged exoplanets, where speckles originating from stellar stray light render the extraction of the continuum difficult. Our framework allows the atmospheric characterisation of directly imaged exoplanets using the high-quality spectral data that will be provided by the new generation of instruments such as the Enhanced Resolution Imager and Spectrograph (ERIS) at the Very Large Telescope, the Mid-Infrared Instrument (MIRI) aboard the James Webb Space Telescope, and in the future the Mid-infrared ELT Imager and Spectrograph (METIS) at the Extremely Large Telescope.}


   \maketitle

\section{Introduction}
\label{sec:introduction}

Determining the ratios of elemental abundances in the atmosphere of exoplanets is emerging as a key analysis to unwrap the history of their formation. Nascent planets do indeed inherit some of the chemical properties of the protoplanetary disk at the location of their formation: \citet{Oberg2011THEATMOSPHERES} showed that the molecular snowlines are expected to directly affect the atmospheric carbon-to-oxygen (C/O) number ratio. Further processes that modify the chemical composition of the disk can add nuance to this initial picture; for example, the inward drift of pebbles covered with CO ice might enhance the abundance of CO gas within the CO iceline~\citep{Oberg2016ExcessGas}, or dust transport processes such as grain growth, settling, and vertical mixing can affect the C/O ratio depending on the location of snowlines and pressure bumps in the disk~\citep{vanderMarel2021IfIt}. Furthermore, the chemical composition of a planet can drastically change after formation due to the later accretion of surrounding material while it migrates. For instance, the accretion of ice-rich planetesimals can lead to the enrichment of metals in the atmosphere of hot Jupiters~\citep{Mordasini2016THEJUPITERS}, or affect the initial atmospheric C/O ratio inherited from core accretion or gravitational collapse and its initial location with respect to the water and CO$_{2}$ icelines~\citep{Nowak2020PeeringInterferometry}. Currently, the details of gas giant formation are still mostly unconstrained. Many interconnected physical processes are believed to influence their final bulk and atmospheric composition. This creates model degeneracies that prevent us from unequivocally inverting their full formation history~\citep{Molliere2022InterpretingAssumptions}. Nevertheless, on a case-by-case basis, measurements of atmospheric composition might be leveraged to identify specific aspects of formation processes; for example, a high N/O ratio may indicate a large orbital distance during formation \citep{Piso2016THEDISKS,Ohno2022NitrogenSpectra}. Meanwhile population analyses might identify statistical trends to inform formation models \citep{Hoch2022AssessingMass}.

It is particularly important to develop robust statistical frameworks in light of the new generation of telescopes and instruments that have recently started operating or are coming up this decade: the James Webb Space Telescope (JWST) with its Mid-Infrared Instrument \citep[MIRI;][]{Wells2015TheSpectrometer}, capable of integral field spectroscopy at a spectral resolution of $R\,\sim$\,\numrange{1500}{3500} between 4.9 and \SI{28.3}{\micro\meter}; the new Enhanced Resolution Imager and Spectrograph \citep[ERIS;][]{Davies2018ERIS:VLT} at the Very Large Telescope (VLT), which contains the newly refurbished Integral Field Unit (IFU) SPIFFIER~\citep{George2016MakingRe-commissioning}, upgraded from the SPectrometer for Infrared Faint Field Imaging (SPIFFI) within the old Spectrograph for INtegral Field Observations in the Near Infrared (SINFONI), with a spectral resolution of $R\,\approx$\,\numrange{5000}{10000} in the J, H, and K bands; and finally the Mid-infrared ELT Imager and Spectrograph~\citep[METIS;][]{Brandl2021METIS} on the upcoming Extremely Large Telescope (ELT), which allows long-slit spectroscopy at low and medium resolution ($R\,\approx$\,\numrange{400}{1900}), as well as integral field spectroscopy at high resolution ($R\,\approx$\,\num{100000}) in the L and M bands.

The first measurement of the atmospheric C/O ratio of an exoplanet was performed by \citet{Madhusudhan2011AWASP-12b}, who applied an atmospheric fitting code to Spitzer and ground-based photometric data of the transiting hot Jupiter WASP-12b. Since then, many atmospheric retrieval codes have been developed for photometric and spectroscopic data of both transiting and directly imaged exoplanets; for example, readers can refer to \texttt{CHIMERA}~\citep{Line2013ATECHNIQUES}, \texttt{NEMESIS}~\citep{Lee2012OptimalSpectroscopy}, \texttt{Tau-REx}~\citep{Waldmann2015TAU-REXATMOSPHERES}, \texttt{HELIOS-RETRIEVAL}~\citep{Lavie2017ttHELIOSRETRIEVAL:/ttFormation}, and \texttt{petitRADTRANS}~\citep{Molliere2019PetitRADTRANS:Retrieval}. On the direct-imaging side, the first proposed strategy to measure the atmospheric C/O ratio has been the combination of photometric and low-resolution ($R<1000$) spectroscopic data: for example, \citet{Lavie2017ttHELIOSRETRIEVAL:/ttFormation} analysed HR~8799 data from \citet{Bonnefoy2016FirstSPHERE} and \citet{Zurlo2016FirstSPHERE} with the VLT Spectro-Polarimetric High-contrast Exoplanet REsearch instrument~\citep[SPHERE;][]{Beuzit2019SPHERE:Telescope} and the Gemini Planet Imager~\citep[GPI;][]{Macintosh2014FirstImager}; \citet{Molliere2020Retrieving8799e} studied HR~8799~e more closely with additional VLTI/GRAVITY spectro-interferometric data; and \citet{Nowak2020PeeringInterferometry} observed $\beta$~Pictoris~b with VLTI/GRAVITY to measure its atmospheric C/O ratio. More recently, \citet{Ruffio2021DeepSpectroscopy} observed the HR~8799 system with the OH-Suppressing Infrared Integral Field Spectrograph ~\citep[OSIRIS;][]{Quirrenbach2003Integral-fieldKeck,Larkin2006OSIRIS:Keck} at the Keck Observatory in the K band at medium spectral resolution ($R \approx 4000$) and measured stellar C/O ratios for the b, c, and d~planets, contradicting the previous findings of \cite{Lavie2017ttHELIOSRETRIEVAL:/ttFormation} of the supersolar C/O ratio for the b planet, hinting at a possible discrepancy between measurements at low versus medium spectral resolution. Finally, \citet{Xuan2022ASpectroscopy} simultaneously analysed photometric and low- and high-resolution spectroscopic data of the brown dwarf HD~4747~B to derive its C/O ratio, while \citet{Wang2023RetrievingData} additionally considered medium-resolution spectroscopy while applying the same framework based on \citet{Wang2020OnC} to constrain the atmospheric parameters of HR~8799~c.

Several studies have explored how to leverage medium-resolution spectroscopy to investigate the atmospheric composition of directly imaged gas giants, in particular when using cross-correlation with spectral templates --- a method informally called cross-correlation spectroscopy (CCS). Relying on it, \citet{Hoeijmakers2018Medium-resolutionImaging} introduced the technique of molecular mapping, which consists in applying a spectral cross-correlation to the spectrum in each spatial pixel (spaxel) of the datacube of an integral field unit (IFU), thereby unveiling the imprint of each molecule in the atmosphere of the observed exoplanet. Instead of filtering out the starlight as part of data post-processing, \cite{Ruffio2019RadialSpectroscopy} --- later improved by \citet{Ruffio2021DeepSpectroscopy} --- defined a joint likelihood function of the forward model for the planetary atmosphere and stellar point spread function (PSF), allowing to constrain the barycentric radial velocity of the exoplanet as well as the atmospheric parameters using a grid of BT-Settl models \citep{Allard2013ThePlanets} after marginalising over the stellar PSF model.

The molecular mapping technique resulted in the detection of water and carbon monoxide in the atmospheres of $\beta$~Pic~b \citep{Hoeijmakers2018Medium-resolutionImaging} and HIP~65426~b \citet{Petrus2021Medium-resolutionB} using K-band medium-resolution ($R \approx 5000$) spectroscopic data obtained with VLT/SINFONI \citep{Eisenhauer2003SINFONIVLT}, as well as in the atmosphere of HR~8799~b \citep{PetitditdelaRoche2018MoleculeKeck} using H- and K-band medium-resolution ($R \approx 4000$) spectroscopic data obtained with Keck/OSIRIS. \citet{Cugno2021MolecularSystem} applied the technique to K-band spectroscopic data of the forming planet PDS~70~b taken with VLT/SINFONI, and quantified the extinction of the surrounding dusty environment required to explain the non-detection of H$_{2}$O and CO. \citet{Patapis2022DirectMIRI} examined the feasibility and limiting factors of applying molecular mapping with the Mid-Infrared Instrument \citep[MIRI;][]{Wells2015TheSpectrometer} on the James Webb Space Telescope (JWST) using mock observations of the systems HR~8799 and GJ~504, and predicted the detectability of H$_{2}$O, CO, CH$_{4}$, and NH$_{3}$ in the atmospheres of directly imaged gas giants.
%
Using the joint modelling of the stellar PSF and exoplanet atmosphere, \citet{Ruffio2019RadialSpectroscopy} inferred the barycentric radial velocity (RV) of the planets HR~8799~b and c using H- and K-band medium-resolution ($R \approx 4000$) spectroscopic data from Keck/OSIRIS, while \citet{Ruffio2021DeepSpectroscopy} constrained the C/O ratios of the HR~8799~b, c, and d~planets.

Arguably, the frameworks described in these studies are missing at least one of two key ingredients, namely 1. the simultaneous consideration of archival photometric and low-resolution spectroscopic data on top of the newly published medium-resolution spectroscopic data within the Bayesian framework, or 2. the commitment to a full-on atmospheric retrieval by exploring the whole parameter space using dedicated sampling algorithms to not only find the maximum likelihood estimator, but also derive the full posterior distribution. 

With this study, we propose a new framework to combine low resolution spectrophotometric and cross-correlation spectroscopic data into a full Bayesian statistical framework to derive atmospheric properties of directly imaged exoplanets and to prepare for newly commissioned and future instruments. To that end, we built upon the work of \citet{Brogi2019RetrievingSpectroscopy}, who defined a Bayesian statistical framework combining low- and high-resolution spectroscopy with the original purpose of deriving atmospheric properties of transiting exoplanets. We present our framework \texttt{CROCODILE} (\textbf{CRO}ss-\textbf{CO}rrelation retrievals of \textbf{D}irectly \textbf{I}maged self-\textbf{L}uminous \textbf{E}xoplanets)\footnote{\texttt{CROCODILE} is available as a Python package at \url{https://github.com/JHayoz/CROCODILE}}, which embeds the radiative transfer routine \texttt{petitRADTRANS}~\citep{Molliere2019PetitRADTRANS:Retrieval} into the mathematical framework of \citet{Brogi2019RetrievingSpectroscopy}. 

In Sect.~\ref{sec:CROCODILE}, we present the main parts of \texttt{CROCODILE}, namely the Bayesian estimator, the modification of the log-likelihood function to include the spectral cross-correlation between model and data, and our forward model for the atmosphere of gas giant exoplanets. In Sect.~\ref{sec:tests}, we put \texttt{CROCODILE} to a full battery of tests, including a comparison between a regular atmospheric retrieval running on low-resolution spectroscopy (LRS) and photometry and a pure cross-correlation-based retrieval, as well as a review of the performance of \texttt{CROCODILE} when applied to a large portion of the parameter space of the atmospheres of gas giants. Finally, we discuss the implications of our findings as well as future opportunities for \texttt{CROCODILE} in Sect.~\ref{sec:discussion}, before summarising our work in Sect.~\ref{sec:conclusion}\looseness=-1

\section{CROCODILE: A Bayesian framework to combine all direct-imaging observing techniques into atmospheric retrievals}
\label{sec:CROCODILE}

\begin{figure}
    \centering
    \includegraphics[width=\hsize]{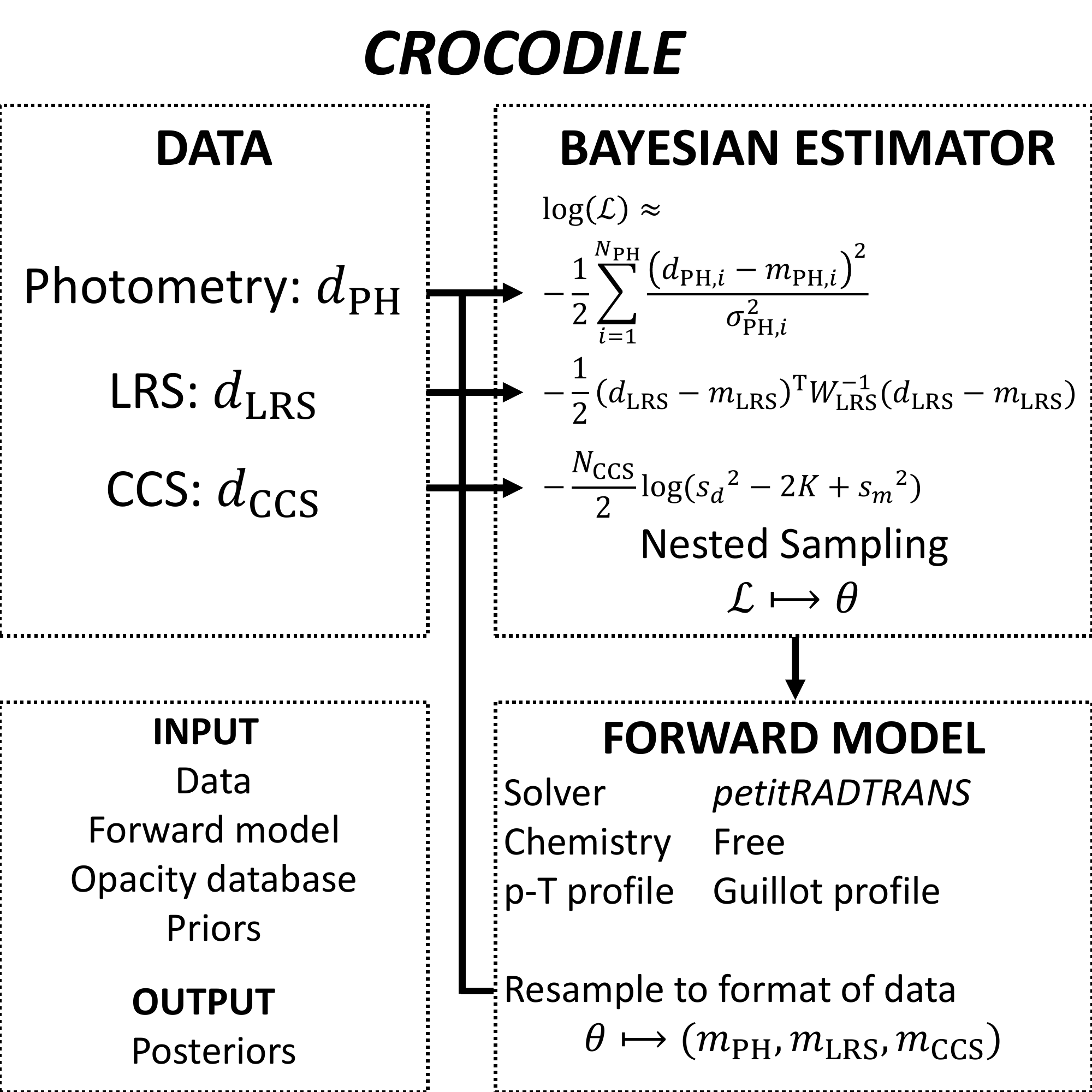}
    \caption{Schematic of our atmospheric retrieval framework \texttt{CROCODILE}. The input of the routine is the data --- which can be photometry, low-, or medium-resolution spectroscopy --- , the forward model, the opacity database, and the priors. \texttt{CROCODILE} starts by sampling parameters from the prior, from which the forward model is evaluated. It is then compared to the data by computing the sum of the log-likelihood functions corresponding to the three observing techniques, from which new parameters are sampled and the loop restarts. The output is the posterior distribution of the forward model.}
    \label{fig:retrieval_framework}
\end{figure}

\texttt{CROCODILE} is composed of a Bayesian estimator coupled to a forward model for the emission spectra of gas giants. More precisely, the routine starts with the prior distributions of the model parameters, from which the Bayesian estimator samples a set of input parameters for the simulator. The forward model computes the spectrum from the sampled parameters, before formatting it to the observing techniques that make up the data, that is photometry, low-resolution spectroscopy, or continuum-removed medium-resolution spectroscopy. These synthetic model spectra are used together with the data to calculate the sum of the three log-likelihood functions associated to each observing technique. Finally, the log-prior is added to the log-likelihood to obtain the log-posterior which is then used to update the sampled points according to a nested sampling algorithm. The input of \texttt{CROCODILE} is the data, the forward model, the opacity database, and the priors. Its output is the posterior distribution of the forward model, from which the posterior thermal structure, chemistry, and spectrum can be computed. A conceptual sketch of \texttt{CROCODILE} is shown in Fig.~\ref{fig:retrieval_framework}, while we describe its implementation in the following sections.

\subsection{The Bayesian estimator}
\label{sec:Bayes}

Atmospheric retrievals are a well-established framework within the exoplanet community \citep[for a review, readers can refer to][]{Madhusudhan2018AtmosphericExoplanets}. Fundamentally, they seek to invert the data, meaning that, for a given model describing the transfer of electromagnetic radiation through the atmosphere of exoplanets, they estimate how well each combination of parameters of said model explains the data. Essentially, they contain three components, namely 1.~the data (which is usually obtained via photometry or spectroscopy), 2.~the forward model (which maps the parameters of the exoplanetary atmosphere to an emission or transmission spectrum), and 3.~a Bayesian inference algorithm that takes in the data as well as prior knowledge and calculates a posterior probability distribution over the parameter space of the model. Bayes' theorem~\citep{Bayes1763LII.S} provides the mathematical framework by formulating the problem as the computation of the posterior distribution:\looseness=-1

\begin{equation}
\label{equ:bayes_theorem}
    p(\theta \mid d) =\frac{p(d \mid \theta) p(\theta)}{p(d)}\,,
\end{equation}

where the vector $\theta$ denotes the parameters of the forward model (e.g. the equilibrium temperature, surface gravity, or chemical composition), $d$ the data,  $p(d \mid \theta) \equiv \Lagr(\theta)$ the likelihood function, $p(\theta)$ the prior distribution, and $p(d) = \int_{\Theta}p(d\mid\theta)p(\theta)$ the evidence defined as the integral of the likelihood function multiplied by the prior over the parameter space $\Theta$. Calculating the likelihood function is expensive since it requires to evaluate the forward model including the computationally heavy radiative transfer routine. Therefore, we need an efficient parameter sampling method to reduce the number of evaluations of the forward model. Our code \texttt{CROCODILE} makes use of the software \texttt{pymultinest}~\citep{Buchner2014X-rayCatalogue}, which is broadly used in the community of atmospheric retrievals \citep[e.g.][]{Lavie2017ttHELIOSRETRIEVAL:/ttFormation,Molliere2019PetitRADTRANS:Retrieval,Molliere2020Retrieving8799e,Konrad2022LargeLIFE}, a Python implementation of \texttt{MultiNest}~\citep{Feroz2008MultiNest:Physics,Feroz2019ImportanceAlgorithm} which utilises the multimodal nested sampling algorithm from \citet{Skilling2006NestedComputation}.

The choice of the likelihood function $\Lagr(\theta)$ depends on the noise statistics of the data, which depend on the observing and post-processing techniques used. Typically, photometric data $d_{\mathrm{PH}}$ obtained by different instruments or in different observing conditions are statistically independent of each other, allowing the decomposition of the likelihood function into a product of likelihood functions for each datapoint. Furthermore, the noise in high-contrast imaging data is usually assumed to be independent identically distributed Gaussian, leading to the following log-likelihood function (ignoring constant factors that only depend on the data):

\begin{equation}
\label{equ:likelihood_photometry}
    \log \Lagr_{\mathrm{PH}}\left(\theta\right) \approx -\frac{1}{2} \sum_{i=1}^{N_{\mathrm{PH}}} \frac{\left( d_{\mathrm{PH},i}-m_{\mathrm{PH},i}\left(\theta\right)\right)^{2}}{\sigma_{\mathrm{PH},i}^{2}}\,,
\end{equation}

where $m_{\mathrm{PH}}(\theta)$ is the synthetic photometric data obtained by evaluating the forward model at the parameter $\theta$, and $\sigma_{\mathrm{PH}}$ is the uncertainty associated to the data $d_{\mathrm{PH}}$.

For low-resolution spectroscopic data $d_{\mathrm{LRS}}$, the noise statistics can be more intricate, since different wavelength channels of a spectrum might be correlated \citep[e.g.][]{Todorov2016THESPECTRUM, Madhusudhan2018AtmosphericExoplanets, Molliere2019PetitRADTRANS:Retrieval, Nowak2020PeeringInterferometry}. In this case, the likelihood function is often chosen as a multivariate normal distribution (again ignoring constant factors):

\begin{equation}
\label{equ:likelihood_gravity}
    \log \Lagr_{\mathrm{LRS}}\left(\theta\right) \approx -\frac{1}{2} \left(d_{\mathrm{LRS}}-m_{\mathrm{LRS}}\left(\theta\right)\right)^{T}W_{\mathrm{LRS}}^{-1}\left(d_{\mathrm{LRS}}-m_{\mathrm{LRS}}\left(\theta\right)\right)\,,
\end{equation}

where $m_{\mathrm{LRS}}(\theta)$ is the synthetic low-resolution spectrum that is obtained by evaluating the forward model at the parameter $\theta$ \citep{Greco2016THEEXOPLANETS}. The covariance matrix $W_{\mathrm{LRS}}$ describes the correlation of the data across spectral channels and needs to be computed for each observation \citep[e.g.][]{Nowak2020PeeringInterferometry}.

\subsection{Incorporating cross-correlation spectroscopy into atmospheric retrievals}
\label{sec:cross_correlation_spectroscopy}

Cross-correlation spectroscopy \citep[CCS;][]{Sparks2002ImagingDetection} was introduced to the field of exoplanet direct-imaging to leverage the full spectral resolution of spectrographs.
With the exception of M stars, the high temperatures found in the photosphere of stars prevent the atoms from bonding into molecules. For that reason, stellar spectra only show atomic absorption lines in the infrared range, which differ significantly from the molecular absorption lines found in colder objects such as brown dwarfs and gas giant exoplanets. Therefore, it is possible to filter out the stellar spectrum and cross-correlate the residuals with molecular spectral templates, thereby picking up the signal of molecules present in the atmosphere of the low-mass companion, for example using high-resolution spectral differential imaging~\citep[HRSDI; e.g.][]{Hoeijmakers2018Medium-resolutionImaging, Haffert2019Two70, Cugno2021MolecularSystem}.

To incorporate HRSDI-treated IFU data into atmospheric retrievals, we followed the framework of \citet{Brogi2019RetrievingSpectroscopy}, who expanded the conventional Gaussian likelihood function from Eq.~\ref{equ:likelihood_photometry} by assuming constant noise across all spectroscopic channels and replacing the standard deviation by its maximum likelihood estimator, thereby making the cross-covariance $K$ between model and data appear in the log-likelihood function:

\begin{equation}
\label{equ:likelihood_CCS}
     \log \Lagr_{\mathrm{CCS}}\left(\theta\right) \approx -\frac{N_{\mathrm{CCS}}}{2}\log \left( s_{d}^{2} - 2K\left(\theta\right) + s_{m}^{2}\left(\theta\right)\right)\,,
\end{equation}

with the following definitions

\begin{align}
    s_{d}^{2} &= \frac{1}{N_{\mathrm{CCS}}}\sum_{i=1}^{N_{\mathrm{CCS}}}d_{\mathrm{CCS},i}^{2} \label{equ:sd2_app} \\
    s_{m}^{2}\left(\theta\right) &= \frac{1}{N_{\mathrm{CCS}}}\sum_{i=1}^{N_{\mathrm{CCS}}}m_{\mathrm{CCS},i}^{2}\left(\theta\right) \label{equ:sm2_app} \\
    K\left(\theta\right) &= \frac{1}{N_{\mathrm{CCS}}} \sum_{i=1}^{N_{\mathrm{CCS}}} d_{\mathrm{CCS},i}m_{\mathrm{CCS},i}\left(\theta\right)\,. \label{equ:K_app}
\end{align}

In Eq.~\ref{equ:K_app}, the cross-covariance $K$ is linked to the cross-correlation $C$ over the relation:

\begin{equation}
\label{equ:K_C}
     K \equiv C\sqrt{s_{d}^{2}s_{m}^{2}}\,,
\end{equation}

meaning that either the cross-correlation $C$ or the cross-covariance $K$ can be computed as long as the right variable is used in Eq.~\ref{equ:likelihood_CCS}. For the sake of completeness, we provide the derivation of Eq.~\ref{equ:likelihood_CCS} used by \citet{Brogi2019RetrievingSpectroscopy} in appendix~\ref{app:derivation_CCS}. So far, we have kept the explicit dependency of the model M on the parameter $\theta$ for the sake of mathematical rigorousness; however, we have dropped the dependency of the model spectrum on the frame of reference. Whereas the effect is insignificant for photometry and low-resolution spectroscopy of close-by stellar systems, the Doppler shift created by the radial velocity $v_\mathrm{R}$ of the target exoplanet with respect to the observatory indeed becomes significant at medium spectral resolution. More precisely, for a spectrum at spectral resolution $R$ and at wavelength $\lambda$, and therefore with channel spacing $\Delta\lambda = \frac{\lambda}{R}$, the Doppler shift $\frac{\Delta\lambda}{\lambda} = \frac{v_{\mathrm{R}}}{c}$ (with $c$ the speed of light in vacuum) resulting from the radial velocity $v_{\mathrm{R}}$ is equal to the wavelength spacing of the spectrum when $v_{\mathrm{R}} = \frac{c}{R}$. At medium spectral resolution ($R\approx5000$), this effect occurs with a radial velocity of $v_{\mathrm{R}} \approx$ \SI{60}{\kilo\meter\per\second}, whereas it requires a much higher radial velocity ($v_{\mathrm{R}} >$ \SI{600}{\kilo\meter\per\second}) at low spectral resolution ($R<500$). Hence, the forward model for medium- and high-resolution spectroscopy needs to be Doppler-shifted to the reference frame of the observed spectrum. Instead of computing it once and for all before the retrieval or adding it in as a parameter of the model, we incorporated it into the algorithm by taking the maximum of the cross-covariance $K$ over a range of radial velocity

\begin{equation}
\label{equ:K_max_vr}
    K\left(\theta\right) = \max_{v_{0} \leq v_{\mathrm{R}} \leq v_{1}} K\left(\theta,v_{\mathrm{R}}\right) \,,
\end{equation}

where $K\left(\theta,v_{\mathrm{R}}\right)$ is the cross-covariance where the model is Doppler shifted by the radial velocity $v_{\mathrm{R}}$, and $v_{0}$ and $v_{1}$ are the lower and higher bounds of the range of radial velocities considered. To Doppler shift the forward model spectrum, we used the methods \texttt{dopplerShift} and \texttt{crosscorrRV} from the Python package \texttt{PyAstronomy}~\citep{PyAstronomy}, and we selected radial velocities between $v_{0}=-400$ and $v_{1}=$~\SI{400}{\kilo\meter\per\second} in steps of \SI{0.5}{\kilo\meter\per\second}, which corresponds to 0.8\% of the wavelength spacing at medium spectral resolution $R=5000$. The size of the steps have conservatively been chosen small because Doppler-shifting the spectra is a cheap computational operation relative to the cost of computing the forward model. 

These algorithms use interpolation to Doppler shift spectra, which can lead to non-conservation of the total flux. To limit this effect, we applied the Doppler shift before down-sampling the spectra to a lower spectral resolution when simulating data. On the other hand, the routine \texttt{crosscorrRV} interpolates the already down-sampled forward model to the Doppler shifted wavelength before computing the correlation. To understand how the use of interpolation might negatively affect the cross-correlation, we implemented and tested an algorithm which first applies the Doppler shift before down-sampling and computing the correlation. To quantify the improvement, we simulated three spectra using our forward model from Sect.~\ref{sec:preliminary_test}: one data spectrum with added random scatter, and two template spectra with different model parameters. We then computed the likelihood ratio (Eq.~\ref{equ:likelihood_CCS}) between the two template spectra using both algorithms. The new algorithm did not significantly change the results: the likelihood ratios differed by $0.03\%$. However, the new algorithm was slower by a factor of 1000 due to the extra down-sampling step, which would become a bottleneck during retrievals. Hence, we did not use our new algorithm in this work and continued evaluating Eq.~\ref{equ:likelihood_CCS} using the routine \texttt{crosscorrRV}. We note that the issue discussed here might yield a different result with another combination of spectral resolution and wavelength range. Therefore, we recommend caution when using an interpolation-based Doppler shifting algorithm; in particular, the test described above should be repeated when working with alternate datasets.

Finally, it is worth stating explicitly that the spectral data $d_{\mathrm{CCS}}$ is assumed to have been reduced using HRSDI, and thus its continuum has been removed. Therefore, the model also needs to be high-pass filtered with the same cutoff frequency as the data. We describe our high-pass filter in Sect.~\ref{section:sim_data}.

\subsection{Combining data from different instruments}
\label{sec:combine_data}

In the two previous sections, we motivated our choice of log-likelihood functions associated to photometric, low-resolution spectroscopic, and cross-correlation spectroscopic data. To integrate all three techniques into one framework, we need to compute their joint likelihood function:

\begin{align}
    \Lagr\left(\theta\right) = p(d_{\mathrm{PH}},d_{\mathrm{LRS}},d_{\mathrm{CCS}}\mid \theta)\,.
\end{align}

In general, the joint likelihood functions of two datasets $d_{1}$ and $d_{2}$ depends on their conditional probability $p(d_{1},d_{2}\mid\theta) = p(d_{1}\mid d_{2}\theta) p(d_{2}\mid\theta)$. In this work, we assumed that all datasets are independent of each other, that is $p(d_{1}\mid d_{2}\theta) = p(d_{1}\mid\theta)$. This assumption should hold whenever two datasets come from different telescopes or different instruments, or if they were taken on different nights or under different meteorological conditions. Therefore, we can apply this argument recursively to the joint likelihood function given above to obtain 

\begin{align}
    p(d_{\mathrm{PH}},d_{\mathrm{LRS}},d_{\mathrm{CCS}}\mid \theta) = p(d_{\mathrm{PH}}\mid \theta)p(d_{\mathrm{LRS}}\mid \theta)p(d_{\mathrm{CCS}}\mid \theta) \,,
\end{align}

which translates to the following log-likelihood function for \texttt{CROCODILE}:

\begin{align}
    \log \Lagr\left(\theta\right) = &\log \Lagr_{\mathrm{PH}}\left(\theta\right) + \log \Lagr_{\mathrm{LRS}}\left(\theta\right) + \log \Lagr_{\mathrm{CCS}}\left(\theta\right) \notag\\
     \approx &-\frac{1}{2} \sum_{i=1}^{N_{\mathrm{PH}}} \frac{\left( d_{\mathrm{PH},i}-m_{\mathrm{PH},i}\left(\theta\right)\right)^{2}}{\sigma_{\mathrm{PH},i}^{2}} \notag\\
      &- \frac{1}{2} \left(d_{\mathrm{LRS}}-m_{\mathrm{LRS}}\left(\theta\right)\right)^{T}W_{\mathrm{LRS}}^{-1}\left(d_{\mathrm{LRS}}-m_{\mathrm{LRS}}\left(\theta\right)\right) \notag\\
       &- \frac{N_{\mathrm{CCS}}}{2}\log \left( s_{d}^{2} - 2K\left(\theta\right) + s_{m}^{2}\left(\theta\right)\right)\,.
\end{align}

\subsection{The forward model}
\label{sec:FM}

The forward model of an atmospheric retrieval is composed of two parts, namely the theoretical model describing the chemical and thermal properties of the planetary atmosphere, and a radiative transfer routine that calculates the spectrum escaping from it. For the latter, we utilised \texttt{petitRADTRANS}~\citep{Molliere2019PetitRADTRANS:Retrieval}, which solves the radiative transfer equation on a discrete 1D plane-parallel atmosphere. It can be used in two spectral modes, namely the high-resolution line-by-line mode (lbl; $\lambda /\Delta\lambda = 10^{6}$) or the medium-resolution correlated-k (c-k) approximation ($\lambda /\Delta\lambda = 10^{3}$). We used the lbl mode when calculating spectra of instruments with a spectral resolution higher than 1000, and the other for lower resolution or photometry.
We set the number of atmospheric layers on which the radiative transfer equation is solved to 100 between a top and bottom pressure of $10^{-6}$ and \SI{1e2}{\bar}.

For the physical model of the atmosphere, two main approaches are used in the literature, namely self-consistent~\citep{Seager2005OnJupiters,Burrows2008TheoreticalData,Fortney2008AAtmospheres,Moses2011DISEQUILIBRIUM209458b,Gandhi2017Genesis:Spectra} and free models \citep{Todorov2016THESPECTRUM,Lavie2017ttHELIOSRETRIEVAL:/ttFormation,Nowak2020PeeringInterferometry,Molliere2020Retrieving8799e}. Self-consistent models aim to account for the known relevant physical processes that might take place in a planetary atmosphere and therefore implement physical constraints such as hydrodynamical, thermodynamical, and chemical equilibria, and sometimes even disequilibria due to dynamical interactions such as, for example, winds~\citep{Gandhi2018RetrievalHyDRA} or photochemistry~\citep{Moses2011DISEQUILIBRIUM209458b}. Evidently, the risk of this approach is to falsely account for the relevant physical processes, for example by wrongly assuming chemical equilibrium~\citep{Madhusudhan2018AtmosphericExoplanets}. The solution to that problem is to leave out physical consistency and adopt a parameterised model that is flexible enough to simultaneously account for both well-known and unknown physical processes. The disadvantage of that approach is that it often requires a higher-dimensional model and that it might return physically impossible solutions. In this work, we selected a mixed approach with a physically motivated analytical thermal structure combined with free chemistry. Indeed, we are mainly interested in constraining the abundances of specific molecules, therefore requiring to fit for each of them independently, whereas investigating the diversity of thermal structures that can be constrained by atmospheric retrievals is beyond the scope of this work.


Specifically, our free chemistry model was parameterised by the mass fractions $X_{i}$ of the molecular abundances which are assumed to be constant vertically across all atmospheric layers \citep[e.g.][]{Todorov2016THESPECTRUM,Line2017UniformDwarfs}. The rest of the atmospheric mass was made of 75\% molecular hydrogen and 25\% helium~\citep{Line2012INFORMATIONLOOK}. We considered the line opacities of H$_{2}$O, CO, CO$_{2}$, CH$_{4}$, H$_{2}$S, FeH, TiO, K, and VO, at low (c-k) and high resolution (lbl) depending on the spectral resolution of the instrument for which the spectrum was calculated. We included Rayleigh scattering of molecular hydrogen and helium together with the Collision-Induced Absorption (CIA) cross-sections of H$_{2}$-H$_{2}$ and H$_{2}$-He. The line lists and opacities used in this study were downloaded at the online documentation of \texttt{petitRADTRANS}\footnote{The documentation of \texttt{petitRADTRANS} is available at \url{https://petitRADTRANS.readthedocs.io/en/latest/content/available_opacities.html}} and are listed in Table~\ref{tab:line_list} for reference.

\begin{table}[t]
    \centering
    \caption{Line lists and opacities included.}
    \begin{tabular}{l l l}
        \toprule
        Opacity      & c-k       & lbl \\
        \midrule
        H2           & ...       & HITRAN  \\
        H$_{2}$O     & HITEMP    & HITEMP   \\
        CO           & HITEMP    & HITEMP   \\
        CH$_{4}$     & ExoMolOP  & YT14     \\
        CO$_{2}$     & ExoMolOP  & HITEMP   \\
        H$_{2}$S     & ExoMolOP  & HITRAN   \\
        FeH          & ExoMolOP  & YT14     \\
        TiO          & ExoMolOP  & ExoMolOP \\
        K            & VALD      & VALD     \\
        VO           & ExoMolOP  & ExoMolOP \\
        \midrule
        Rayleigh     &           &          \\
        H2           & DW62      & DW62     \\
        He           & CD65      & CD65     \\
        \midrule
        Quasi-continuum &        &          \\
        H2-H2 CIA    & BR, RG12  & BR, RG12 \\
        H2-He CIA    & BR, RG12  & BR, RG12 \\
        \bottomrule
    \end{tabular}
    \tablefoot{The acronyms given above stand for the following references: ExoMolOP: \citet{Chubb2021TheAtmospheres}, HITEMP: \citet{Rothman2010HITEMPDatabase}, HITRANS: \citet{Rothman2013TheDatabase}, YT14 = \citet{Yurchenko2014ExoMol1500K}, VALD: \citet{Piskunov1995VALD:Base.}, DW62: \citet{Dalgarno1962RayleighHydrogen.}, CD65: \citet{Chan1965TheHelium}, BR: \citet{Borysow1988Collison-inducedK,Borysow1989Collision-inducedK,Borysow1989Collision-inducedBands,Borysow2001High-temperatureAtmospheres,Borysow2002Collision-induced1000K}, RG12: \citet{Richard2012NewCIA}.}
    \label{tab:line_list}
\end{table}


We parameterised the thermal structure of the atmosphere with the analytic semi-grey model derived by \citet{Guillot2010OnAtmospheres}. It is parameterised by six quantities, of which the equilibrium temperature $T_{\mathrm{equ}}$ and surface gravity $g$ are free parameters of our forward model, whereas the other four are fixed: the average opacity in the infrared $\kappa_{\mathrm{IR}}$ = \SI{0.01}{\centi\meter\squared\per g}, the ratio of the average opacity in the optical and infrared $\gamma=\kappa_{\mathrm{V}}/\kappa_{\mathrm{IR}} = 0.4$, the interior temperature $T_{\mathrm{int}}$ = \SI{200}{\kelvin}, and the bottom pressure $P_{0}$ = \SI{100}{\bar}.

The last parameter was the planet radius $R$ which, together with the distance from the Earth $d$, served to scale the atmospheric spectrum down by a multiplicative factor of $R^{2}/d^{2}$, since \texttt{petitRADTRANS} calculates the flux escaping the top of the atmosphere of the planet per unit surface.\looseness=-1

In this study, we purposefully ignored clouds, which are known to be present in the atmospheres of gas giant exoplanets \citep{Nowak2020PeeringInterferometry,Molliere2020Retrieving8799e}. However, considering clouds would add too much complexity to our model and would dilute our message. Indeed, the modelling of clouds is an issue that all atmospheric retrieval frameworks face and therefore need to be investigated separately. 

Finally, we ignored rotational broadening and limb darkening. Indeed, \citet{Palma-Bifani2023PeeringPic} included both parameters into their retrieval on VLT/SINFONI J-, H-, and K-band spectroscopic observations of AB~Pic~b and reported an erratic behaviour of their posteriors, which was probably due to the insufficient spectral resolution of VLT/SINFONI. Furthermore, there was no noticeable correlation between $vsin(i)$ and any other parameter in their published 2-D corner-plots (figures 7 and 8). Similarly, \citet{Bryan2020ObliquityCompanion} measured the rotational broadening of the planetary-mass companion 2MASS~J01225093–2439505 using high-resolution ($R\approx25000$) spectroscopic observations taken with Keck/NIRSPEC and reported no significant dependence on the C/O ratio of the atmospheric models used to cross-correlate with the data.

\section{Tests on simulated data}
\label{sec:tests}

In the following, we aim to demonstrate the capability of \texttt{CROCODILE} to recover the thermal and chemical properties of the atmospheres of gas giants using a common combination of ground-based techniques in the near and mid infrared such as photometry, low-resolution spectroscopy (LRS), and medium-resolution cross-correlation spectroscopy (MRCCS). We therefore submit it to a full series of tests that we describe and justify in the following.

As a preliminary test of \texttt{CROCODILE}, we checked our statistical framework in Sect.~\ref{sec:preliminary_test} for potential mistakes or biases. 
For this, we controlled our implementation of the cross-correlation-based likelihood given by Eq.~\ref{equ:likelihood_CCS} by running a retrieval with MRCCS data alone. 
Then, we tested our implementation of the regular retrieval framework on LRS and photometric (LRSP) data. 
Finally, we tested the implementation of \texttt{CROCODILE} on the combination of the three techniques.\looseness=-1

To test \texttt{CROCODILE} in a meaningful way, we considered a realistic study case for ground-based instruments. This had the following two consequences, namely first that we chose a well-known target to simulate for which some atmospheric properties have been derived already, and second we assumed a realistic list of instruments to simulate spectroscopic and photometric observations with. To that end, we selected $\beta$~Pictoris~b \citep{Lagrange2010APictoris} as a bright and young exoplanet that was observed by many instruments and was the object of several atmospheric studies \citep[e.g.][]{Hoeijmakers2018Medium-resolutionImaging,Stolker2020MIRACLES:B,Nowak2020PeeringInterferometry}. We describe this target and the used instruments in detail in Sect.~\ref{section:sim_target}~and~\ref{section:sim_data}, and show the results of that test in Sect.~\ref{sec:combine_MRCCS_LRS_PHOT}.

Furthermore, we show that combining MRCCS with LRSP generally allows to better constrain the forward model as compared to using only LRSP data as it has been done until now, by applying our framework to different values of the parameter space of the atmospheres of gas giants (readers can refer to Sect.~\ref{section:sim_target} for details).
The results of that test are presented in Sect.~\ref{sec:retrieval_study}.

\subsection{Simulated targets}
\label{section:sim_target}

To test our method, we need to know the ground truth values of the parameters of an exoplanetary atmosphere and see if we can retrieve them truthfully. 
However, it would not be realistic (namely: too optimistic) to use the same chemical model to simulate such an atmosphere that we use as our forward model.
In particular, the underlying chemical model of our target should have a more complex chemistry than our simple forward model. 
Therefore, whereas our forward model assumed free chemistry as we described in Sect.~\ref{sec:FM}, for our simulated targets we chose the chemical equilibrium model that was computed by \texttt{easyCHEM} from \citet{Molliere2017ObservingObservations}\footnote{A grid of molecular abundances pre-computed by \texttt{easyCHEM} can be found at \url{https://petitradtrans.readthedocs.io/en/latest/content/notebooks/poor_man.html}} as a first approximation of the chemistry of an exoplanetary atmosphere, not accounting for clouds \citep{Burrows1999ChemicalAtmospheres}, vertical mixing \citep{Griffith2000Equilibrium229B}, or other processes causing disequilibrium chemistry. 
Concretely, it was calculated in the following way: we set the pressure-temperature profile as well as the carbon-to-oxygen (C/O) ratio and metallicity (Fe/H) and the code calculated the resulting steady-state molecular abundances at chemical equilibrium in each layer of the atmosphere by minimising the Gibbs free energy. We considered the opacities of H$_{2}$O, CO, CH$_{4}$, CO$_{2}$, H$_{2}$S, NH$_{3}$, FeH, HCN, TiO, PH$_{3}$, K, VO, and Na \citep[i.e. HCN in addition to the opacities included in][]{Nowak2020PeeringInterferometry}. We included more opacities here than in our forward model described in Sect.~\ref{sec:FM}. Moreover, we also included Rayleigh scattering of molecular hydrogen and helium together with the Collision-Induced Absorption (CIA) cross-sections of H$_{2}$-H$_{2}$ and H$_{2}$-He. Finally, we used the same p-T profile as our forward model~\citep{Guillot2010OnAtmospheres}. The spectra of our synthetic targets were also generated with \texttt{petitRADTRANS}~\citep{Molliere2019PetitRADTRANS:Retrieval}.\looseness=-1

As mentioned above, we aim to test \texttt{CROCODILE} on a meaningful subspace of exoplanetary atmospheres. In particular, we considered the three parameters that control the chemical equilibrium model, namely the equilibrium temperature, the carbon-to-oxygen ratio, and the metallicity, and we selected a baseline equal to the values derived for $\beta$~Pic~b by \citet{Nowak2020PeeringInterferometry}, namely $T_{\mathrm{equ}}$ = \SI{1742}{\kelvin}, C/O = \num{0.44}, and Fe/H = \SI{0.66}{\dex}, and subsequently varied each of these parameters. Table~\ref{tab:simulated_values} summarises the full list of exoplanetary atmospheres simulated. 
We point out that the values given in the table did not exactly match the input values used for \texttt{easyCHEM} (which were 0.3, 0.44, and 1 for C/O and -1, 0.66, and 2 for Fe/H), since we only considered the gaseous content of the atmosphere and not the liquid and solid reactants available with \texttt{easyCHEM}. The values given in Table~\ref{tab:simulated_values} are the mean C/O ratio and metallicity along the vertical extent of the simulated atmospheres after removing all opacities from non-gaseous reactants.

\begin{table}[t]
    \centering
    \caption{Parameter values assumed for the simulated datasets.}
    \begin{tabular}{c c S[table-format=1.2] S[table-format=3.2]}
        \toprule
         {ID} & {$T_{\mathrm{equ}}$ (K)} & {C/O} & {Fe/H} \\
         \midrule
         1 & 1742 & 0.49 & 0.61 \\
         2 & 1400 & 0.53 & 0.59 \\
         3 & 2000 & 0.47 & 0.62 \\
         4 & 1742 & 0.32 & 0.75 \\
         5 & 1742 & 1.04 & 0.38 \\
         6 & 1742 & 0.46 & -1.02 \\
         7 & 1742 & 0.52 & 1.76 \\
         \bottomrule
    \end{tabular}
    \tablefoot{Each row corresponds to a simulated atmosphere used to compute synthetic MRCCS, LRS, and photometric measurements as described in Sect.~\ref{section:sim_data}.}
    \label{tab:simulated_values}
\end{table}

\begin{table*}[t]
    \centering
    \caption{List of observations of $\beta$ Pic b used for our simulations.}
    \begin{tabular}{l l c c l l}
    \toprule
    Instrument & Band & Technique & Wavelength (\si{\micro\meter}) & $R$ & Reference \\
    \midrule
    VLT/SPHERE/IFS   & Y-H   & LRS   & 0.96--1.63 & 30   & ... \\
    VLTI/GRAVITY     & K     & LRS   & 1.97--2.49 & 500  & \cite{Nowak2020PeeringInterferometry} \\
    VLT/SINFONI      & K     & MRCCS & 2.09--2.45 & 4000 & \cite{Hoeijmakers2018Medium-resolutionImaging} \\
    LCO/VisAO        & Ys    & PHOT  & 0.99       & ...  & \cite{Males2014MagellanNici} \\
    VLT/NACO         & J     & PHOT  & 1.28       & ...  & \cite{Currie2013AB} \\
    Gemini/NICI      & ED286 & PHOT  & 1.59       & ...  & \cite{Males2014MagellanNici} \\
    VLT/NACO         & H     & PHOT  & 1.67       & ...  & \cite{Currie2013AB} \\
    VLT/SPHERE/IRDIS & K12.1 & PHOT  & 2.11       & ...  & \cite{Lagrange2020UnveilingData} \\
    VLT/NACO         & Ks    & PHOT  & 2.15       & ...  & \cite{Bonnefoy2011HighM} \\
    VLT/SPHERE/IRDIS & K12.2 & PHOT  & 2.26       & ...  & \cite{Lagrange2020UnveilingData} \\
    VLT/NACO         & NB374 & PHOT  & 3.74       & ...  & \cite{Stolker2020MIRACLES:B} \\
    VLT/NACO         & Lp    & PHOT  & 3.81       & ...  & \cite{Stolker2019PynPoint:Data} \\
    VLT/NACO         & NB405 & PHOT  & 4.07       & ...  & \cite{Stolker2020MIRACLES:B} \\
    VLT/NACO         & Mp    & PHOT  & 4.79       & ...  & \cite{Stolker2019PynPoint:Data} \\
    \bottomrule
    \end{tabular}
    \tablefoot{The VLT/SPHERE/IFS spectrum is not public and therefore we do not put a reference next to it, however we do refer to \citet{Langlois2021TheSHINE} to estimate the errorbars of our synthetic spectra.}
    \label{tab:observations}
\end{table*}

Additionally, our simulated targets had a planetary radius $R_{\mathrm{p}}$~=~1.36\,R$_{\mathrm{J}}$ \citep{Nowak2020PeeringInterferometry} and an arbitrary radial velocity of $v_{\mathrm{R}}~=~\SI{31}{\kilo\meter\per\second}$, the former being a parameter of the forward model while the latter only plays a role at medium (or higher) spectral resolution and gets determined at every computation of the likelihood function using Eq.~\ref{equ:K_max_vr}. Finally, our model of the targets implicitly included six more parameters which were kept fixed between the data and the forward model. Five of these describe the p-T profile and are the same as given in Sect.~\ref{sec:FM}.
The remaining parameter was the distance $d~=~\SI{19.75}{\parsec}$ to $\beta$~Pictoris~b as measured by \citet{GaiaCollaboration2018VizieR2018}. 

\subsection{Synthetic data}
\label{section:sim_data}

For each of the seven synthetic targets described above, as well as for each evaluation of our forward model, we recreated synthetic photometry, LRS, and MRCCS by matching the instrumental characteristics of the archival datasets of $\beta$ Pic b listed in Table~\ref{tab:observations}. Considering real observations of $\beta$ Pic b enabled us to simulate realistic measurement uncertainties and random noise for each instrument in a straightforward way, thereby creating a realistic synthetic dataset. 
To simulate the format recorded by each of these instruments, we started by computing a spectrum with \texttt{petitRADTRANS} using the atmospheric model described in Sec.~\ref{section:sim_target} in either the full spectral resolution line-by-line mode ($R \approx 10^{6}$) for VLT/SINFONI and VLTI/GRAVITY or the correlated-k approximation ($R\approx1000$) for VLT/SPHERE and photometry.

Photometry was calculated by integrating the spectrum multiplied by the normalised filter transmission function, which we downloaded with the Python package \texttt{species}~\citep{Stolker2020MIRACLES:B}\footnote{\url{https://github.com/tomasstolker/species}}. We estimated the uncertainty of the photometric flux by taking the same relative error as the real measurement, and added normally distributed random scatter with standard deviation equal to the estimated uncertainty.

For the VLTI/GRAVITY instrument, the resampling from the \texttt{petitRADTRANS} to the observation bins was done using the Python package \texttt{spectRES} \citep{Carnall2017SpectRes:Python} which preserves the total flux. We then estimated the uncertainty of the synthetic fluxes by scaling the covariance matrix of the real measurement to the simulated spectrum.
To create realistic flux measurements, we added a random (multivariate normally distributed) scatter to each flux bin with a covariance matrix equal to the simulated covariance matrix. 
For VLT/SPHERE, where we had no access to a spectrum of $\beta$~Pic~b, we chose a simpler approach by setting the uncertainty to $10\%$ (which is an optimistic estimation of the errorbars found in Fig.~11 of \citet{Langlois2021TheSHINE}) and added normally distributed random scatter in the same way as for photometry.

Concerning the VLT/SINFONI MRCCS data, we first applied a Doppler shift with a radial velocity of \SI{31}{\kilo\meter\per\second} to  the spectrum before resampling the spectrum using \texttt{spectRES}. To match the treatment of HRSDI, we applied a low-pass filter consisting of a Gaussian filter with standard deviation of 4.9 nm to compute the continuum, which we subsequently subtracted from the spectrum.
We found the value of 4.9 nm by repeating our preliminary test from Sect.~\ref{sec:preliminary_test} for different filter sizes (readers can refer to App.~\ref{app:filter_size} for details); however, we note that the optimal filter size should be investigated separately when using a real observation.
To add random scatter to our MRCCS data, we aimed to reproduce the signal-to-noise ratio (S/N) of the real observations that we considered. Therefore, we generated a continuum-removed spectrum using the values retrieved by \cite{Nowak2020PeeringInterferometry} for $\beta$ Pic b and cross-correlated it with the SINFONI spectrum from \cite{Hoeijmakers2018Medium-resolutionImaging}. We then took the ratio of the peak of the cross-correlation function to its standard deviation \SI{200}{\kilo\meter\per\second} away from the peak (similar to \citealt{Cugno2021MolecularSystem}) and obtained a value of 14.4. Thus, when we simulated SINFONI spectra, we added Gaussian noise with a standard deviation such that the CCF between noisy and clean spectra also yielded the same ratio of the peak to standard deviation. This made the synthetic SINFONI planet spectra similar (in terms of S/N) to the $\beta$~Pic~b data presented in \cite{Hoeijmakers2018Medium-resolutionImaging}.\looseness=-1

\subsection{Verifying the framework of \texttt{CROCODILE}}
\label{sec:preliminary_test}

To verify the statistical framework used in \texttt{CROCODILE}, we submitted it to a preliminary test in an idealised setup. To that end, we created yet another synthetic target with the same free chemistry as in our forward model. More precisely, we used the same p-T profile as in our simulated atmosphere 1 from Table~\ref{tab:simulated_values} but changed the chemical composition to only contain H$_{2}$O and CO with vertically constant mass fractions equal to \SI{-1.8}{\dex} and \SI{-1.6}{\dex} respectively. Moreover, we did not add any random scatter contained in our low-resolution datasets to ensure that, if the retrieval did not converge to the input values, it was not because of the noise but rather because of a mistake in our framework. Finally, we reduced the amount of random scatter by a factor of three in the MRCCS data with respect to the synthetic observations introduced in Sect.~\ref{section:sim_data} instead of removing it completely because otherwise the argument of the logarithm could become zero if the forward model matches the simulated data.

\begin{figure}[t]
    \centering
    \includegraphics[width=\hsize]{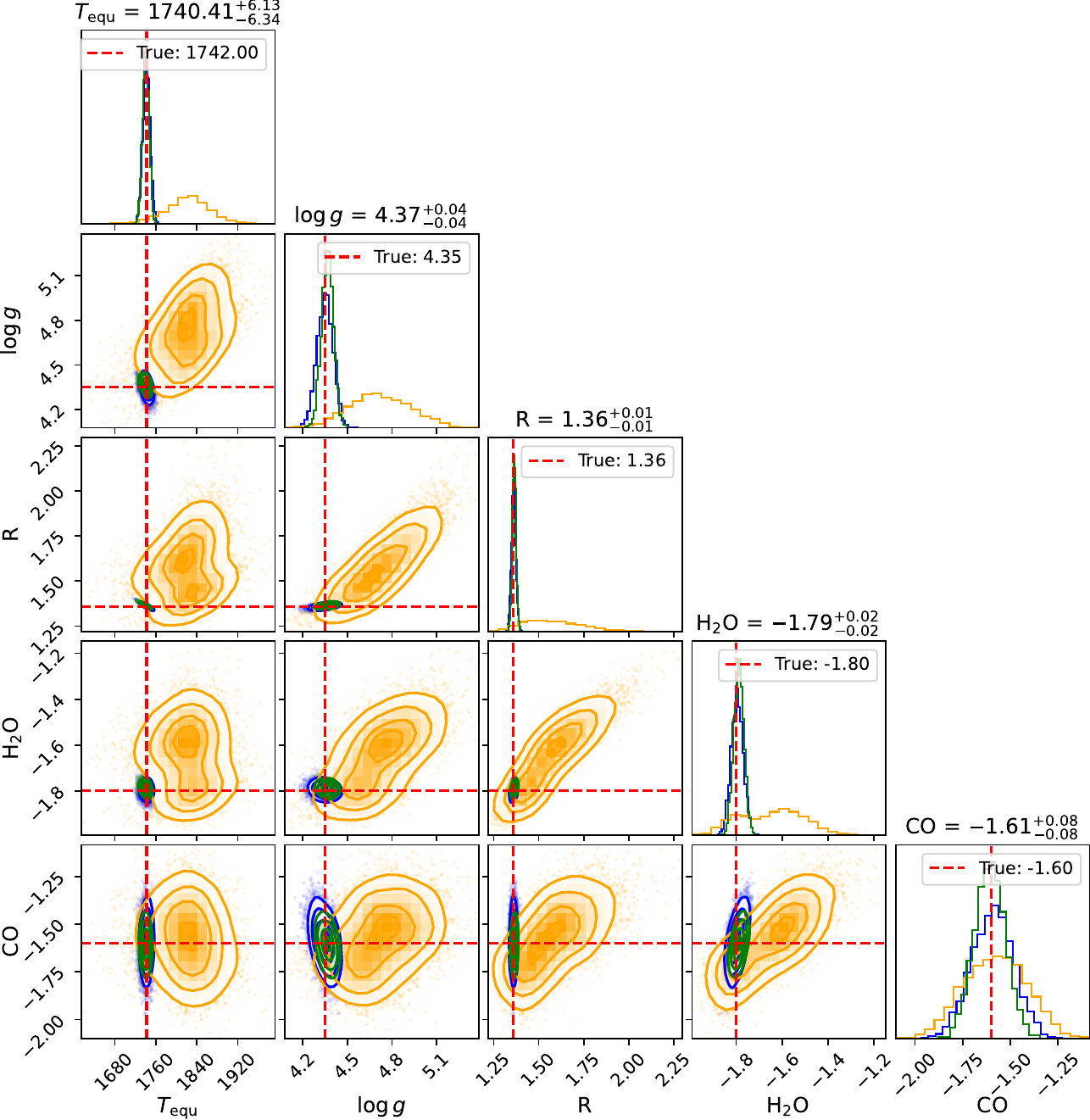}
    \caption{Resulting posterior distributions for our preliminary test that controls our implementation of \texttt{CROCODILE} using the same chemistry as input as our forward model, namely vertically constant abundances of H$_{2}$O and CO, and reduced random scatter. The orange posterior shows the results of the retrieval using only MRCCS data to test the implementation of the cross-correlation likelihood function, the blue posterior is for the regular framework using only LRSP, and the green posterior shows the result for the combination of MRCCS and LRSP data. The dashed red line shows the ground truth used as input.}
    \label{fig:preliminary_test_techniques}
\end{figure}

We first ran \texttt{CROCODILE} on MRCCS data alone to test the implementation of the likelihood function based on the cross-correlation function (Eq.~\ref{equ:likelihood_CCS}). We then tested the implementation of the regular likelihood function based on the residuals between data and model (Eq.~\ref{equ:likelihood_gravity}) using only LRSP. Finally, we checked that the combination of MRCCS together with LRSP via the addition of their log-likelihood functions was also valid. We show the resulting 2D posterior distributions of these three evaluations of \texttt{CROCODILE} in Fig.~\ref{fig:preliminary_test_techniques}.

The three tests yielded educative results: the retrieval on LRSP data alone yielded perfectly centred posteriors with small uncertainty as can be expected from the lack of random scatter in the data. In particular, the abundances of water and carbon monoxide were well constrained due to the good coverage of their spectral features by the VLT/SPHERE/IFS/Y-H and VLTI/GRAVITY/K low-resolution spectroscopic data. On the other hand, the retrieval on MRCCS data yielded a systematic overestimation of the temperature, surface gravity, radius, and H$_{2}$O abundance, while the abundance of CO was constrained very accurately. Furthermore, the radius seemed correlated with the surface gravity and the abundances of H$_{2}$O and CO, and to a lesser extent to $T_{\mathrm{equ}}$.

These correlations can be tentatively explained by the removal of the continuum in the treatment of the MRCCS data in the following way. Without continuum, the radius only has the role of scaling the absorption lines, which is counteracted on the one hand by the surface gravity, which in the p-T model from \citet{Guillot2010OnAtmospheres} brings the same temperatures at higher pressures, which reduces the thermal gradient between the top of the atmosphere and the region at optical depth of unity and hence decreases the strength of the line. The molecular abundances also counteract the scaling effect of the radius by increasing the height at which the optical depth reaches unity, thereby also reducing the thermal gradient between the top of the atmosphere and the emitting layer. Interestingly, if the radius is already known and fixed to a previously measured value during the retrieval, then the results improve greatly as can be seen in Fig.~\ref{fig:preliminary_test_techniques_noR} in the appendix. Indeed, the abundances of H$_{2}$O and CO and the surface gravity were all retrieved within $1\sigma$ with a smaller uncertainty; however, the temperature was still equally overestimated. This test indirectly hints that if the radius --- or any other flux scaling parameter --- is set as fixed parameter to a wrong value, then there is a high risk to bias the results when using MRCCS data only.

Finally, when combining LRSP and MRCCS data, the results were slightly more accurate than when used on LRSP data alone. In particular the abundance of CO was retrieved with a $25\%$ smaller $1\sigma$ spread.
In summary, this test verified the statistical framework of \texttt{CROCODILE} and already showed the systematics that appear when using only MRCCS data.

\subsection{Case study: A synthetic \texorpdfstring{$\beta$}~~Pictoris~b}
\label{sec:combine_MRCCS_LRS_PHOT}

\begin{figure}[t]
    \centering
    \includegraphics[width=\hsize]{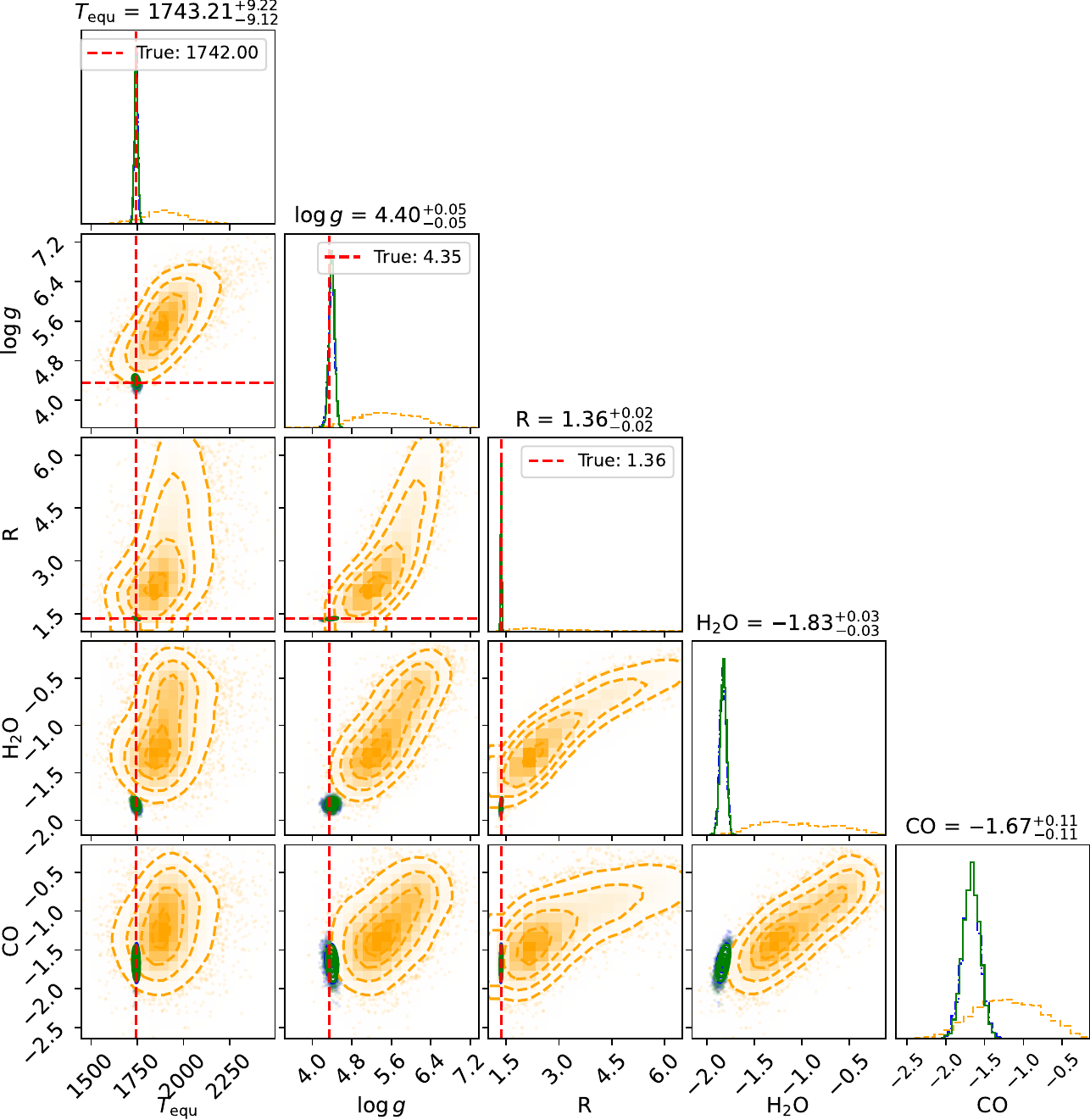}
    \caption{Results of the three retrievals of the same simulated atmosphere using different combinations of the synthetic spectra of $\beta$~Pic~b: the SINFONI MRCCS-only retrieval in orange, the LRSP-only retrieval in blue, and the retrieval using all techniques in green. The off-axis panels show the projected 2D posterior distributions of the main model parameters (readers can refer to the description in Sect.~\ref{sec:FM}). The indicated values above each histogram on the diagonal refer to the median and 1-sigma uncertainty for \texttt{CROCODILE}. For the molecular abundances, we omit the logarithm in front of the names of the molecules for the sake of readability. The dashed red line represents the true parameter value, and it is not shown for the abundances given the different chemical treatment between simulated observations and forward model. }
    \label{fig:doubleRetrieval_small_corner}
\end{figure}

\begin{figure}[!h]
    \centering
    \includegraphics[width=\hsize]{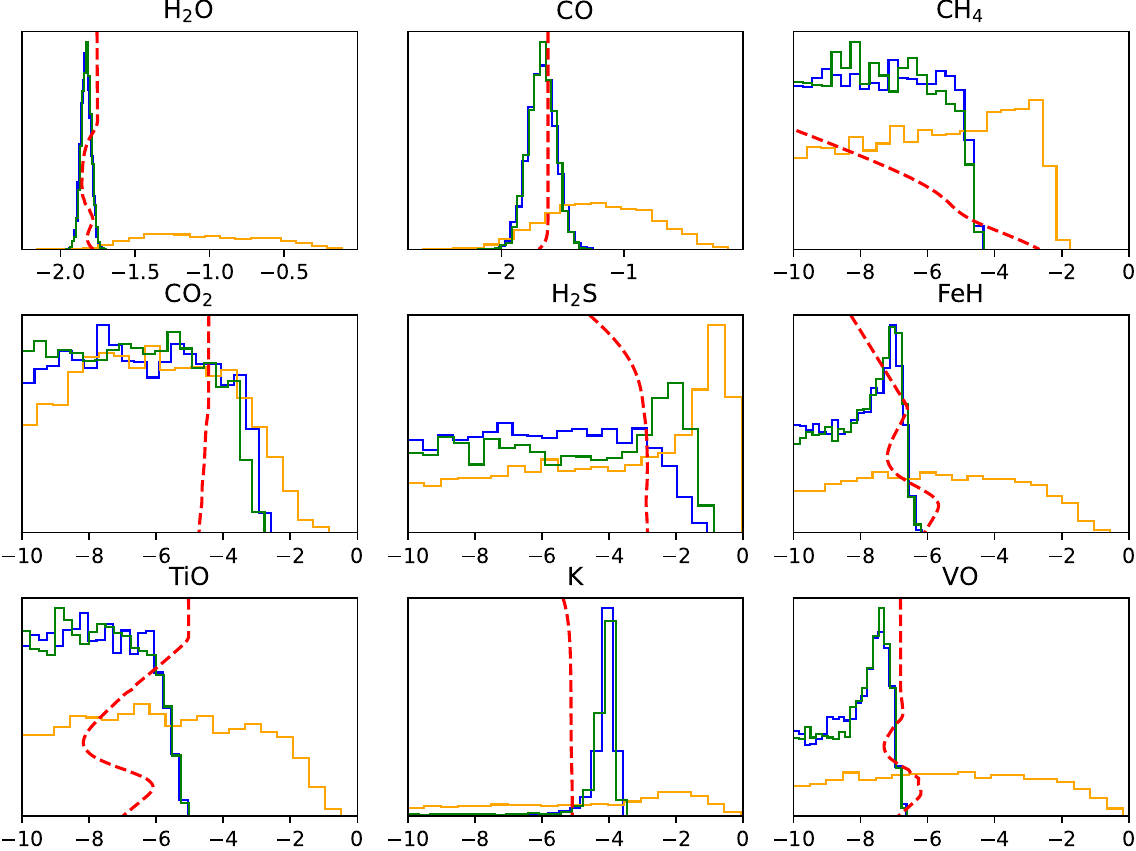}
    \caption{Retrieved posterior distributions of all the molecular abundances included in the three retrievals of the experiment described in \ref{sec:combine_MRCCS_LRS_PHOT}. The solid red lines represent the profile (in log scale) of each molecular abundance along the vertical extent of the atmosphere between \SI{100}{\bar} at the bottom and \SI{1e-6}{\bar} at the top.}
    \label{fig:doubleRetrieval_molecules}
\end{figure}

\begin{figure}[t]
     \centering
     \begin{subfigure}[b]{\hsize}
         \centering
         \includegraphics[width=\textwidth]{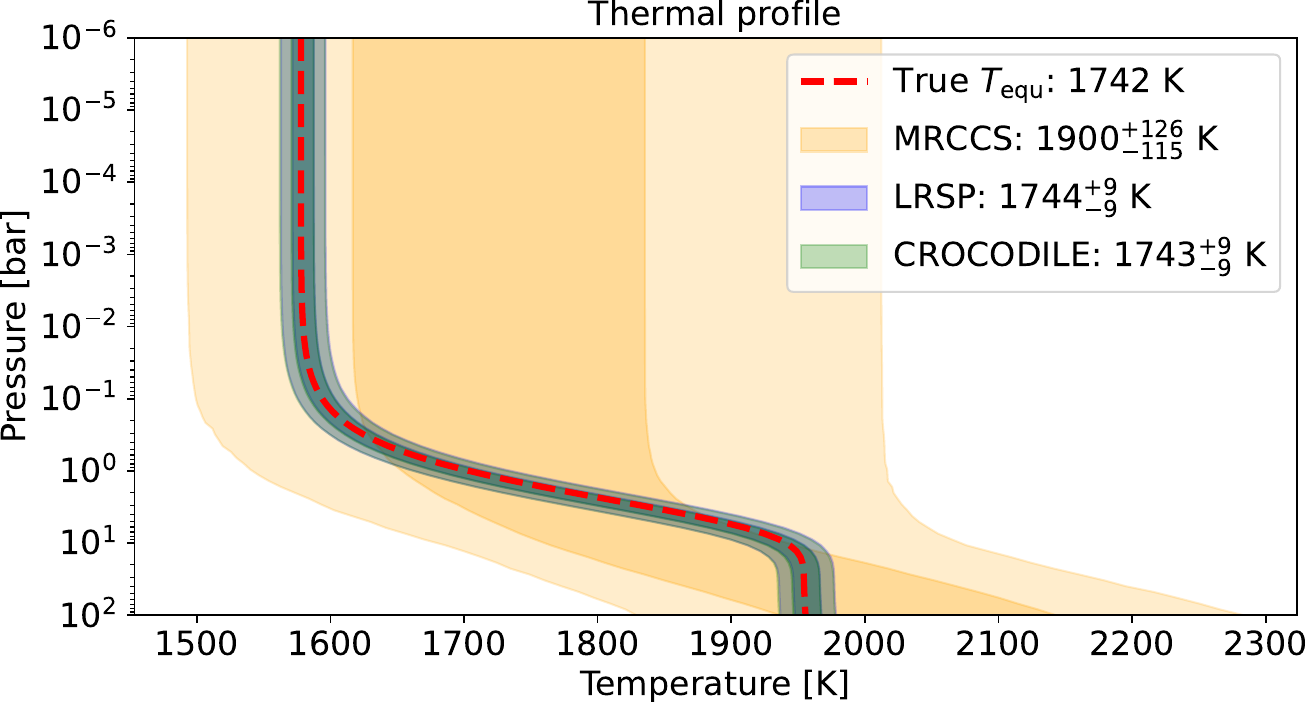}
         \label{fig:study_case_temperature}
     \end{subfigure}
     \hfill
     \begin{subfigure}[b]{\hsize}
         \centering
         \includegraphics[width=\textwidth]{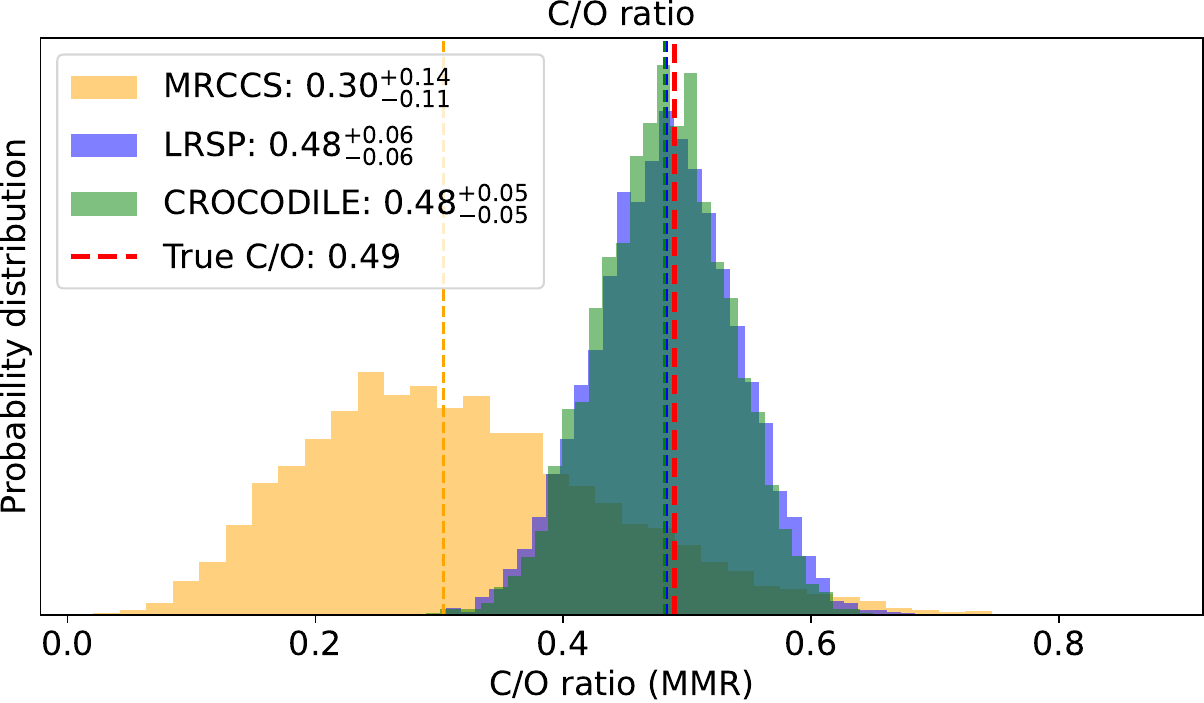}
         \label{fig:study_case_CO}
     \end{subfigure}
     \hfill
     \begin{subfigure}[b]{\hsize}
         \centering
         \includegraphics[width=\textwidth]{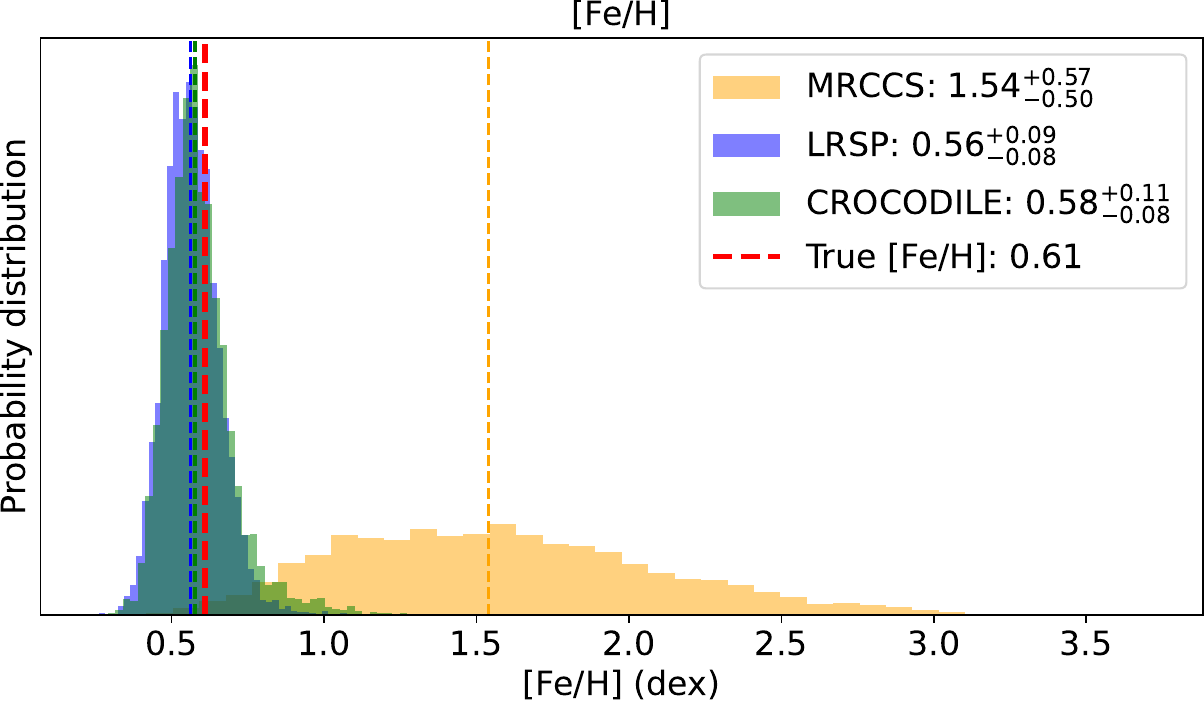}
         \label{fig:study_case_FeH}
     \end{subfigure}
        \caption{Results of the experiment described in \ref{sec:combine_MRCCS_LRS_PHOT}. Top panel: thermal profiles retrieved for each technique (MRCCS: orange, LRSP: blue, \texttt{CROCODILE}: green). The two levels of opacity correspond to the 16 to 84 and the 2 to 98 percentile envelopes of the thermal profile. The legend reports the median and \SI{64}{\percent} confidence interval retrieved for the equilibrium temperature. Middle and bottom panel: posterior distributions of the C/O ratio and metallicity Fe/H inferred from the retrieved abundances for each technique. The dashed lines represent the posterior median. The dashed red line shows the ground truth in every panel.}
        \label{fig:doubleRetrieval_temperature_CO_FeH}
\end{figure}

After validating our statistical framework, we investigated possible biases by applying it to the synthetic targets described in Sect.~\ref{section:sim_target}, which span across a portion of the parameter space around the gas giant $\beta$~Pic~b. First, we give a more in-depth report for our baseline in this subsection while we describe all our results in the next. Figure~\ref{fig:doubleRetrieval_small_corner} compares the posterior distributions obtained with MRCCS data only, with LRSP, or with MRCCS and LRSP simultaneously; the posterior distributions of the molecular abundances are shown in Fig~\ref{fig:doubleRetrieval_molecules}; and the retrieved thermal profiles, C/O ratio, and metallicity Fe/H are presented in Fig.~\ref{fig:doubleRetrieval_temperature_CO_FeH}. Our retrieval setup and priors are described in App.~\ref{app:retrieval_setup}.

The results obtained with MRCCS data only (in orange in Fig.~\ref{fig:doubleRetrieval_small_corner}) showed that the surface gravity, planet radius, and abundances of water and carbon monoxide all yielded Gaussian marginal distributions with the simulated values outside of $1\sigma$ of the mean. The equilibrium temperature was overestimated by $1\sigma$ resulting in a p-T profile that was about \SI{150}{\kelvin} too hot at all altitudes relative to the ground truth, similarly to the planetary radius which was also overestimated. This behaviour is identical to what we observed in the previous section, with similar correlations between the same parameters, once again demonstrating the effects of the removal of the continuum. 
The posterior distributions of the molecular abundances resulted in a good constraint of CO with a large standard deviation of \SI{0.4}{\dex}. The abundance of H$_{2}$O was overestimated by \SI{1}{\dex}, while we obtained correct upper bounds (i.e. the true value was indeed lower than the bound) for CH$_{4}$ and CO$_{2}$, and uniform posteriors for FeH, TiO, H$_{2}$S, K, and VO, which is expected from the weak spectral signatures of these molecules in the K band. The C/O ratio and metallicity did not get retrieved within $1\sigma$, despite large standard deviations of \num{0.14} and \SI{0.5}{\dex} respectively.

\begin{figure*}[!ht]
    \centering
    \includegraphics[width=\hsize]{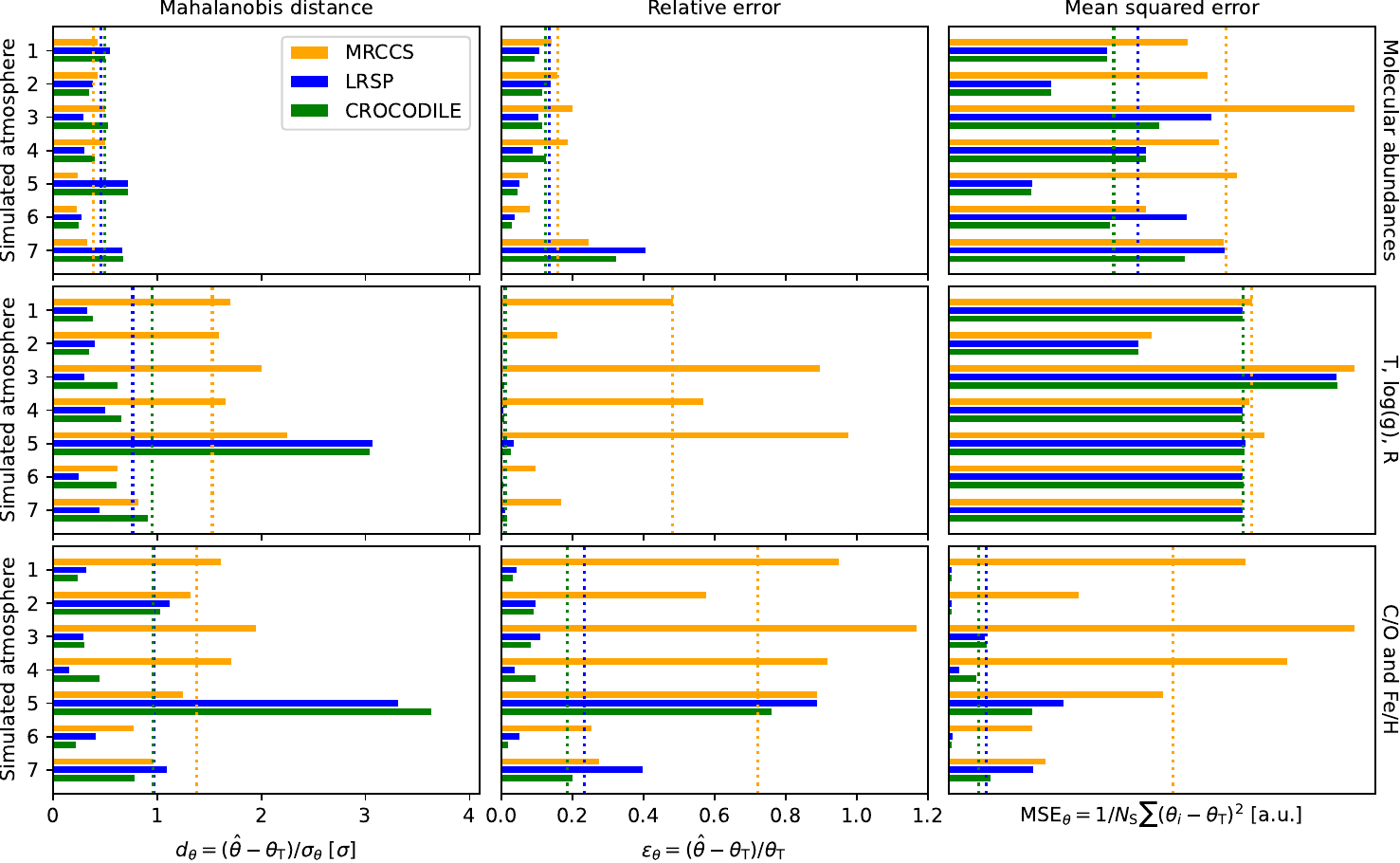}
    \caption{Scores of the three metrics summarising the results of our exploration of the parameter space of gas giants using LRSP, MRCCS, or LRSP+MRCCS (\texttt{CROCODILE}). Each column of panels represents a different metric, and the three rows refer to different parameters grouped together. Within each panel, the score is shown for each technique and for each simulated atmosphere (c.f. Table~\ref{tab:simulated_values} for the input parameters), and averaged over the group of parameters. The dotted lines show the average score over the seven cases.}
    \label{fig:parameter_space_score}
\end{figure*}

When we applied \texttt{CROCODILE} to LRSP data only (in blue in Fig.~\ref{fig:doubleRetrieval_small_corner}), the equilibrium temperature, surface gravity, and planetary radius were retrieved within $1\sigma$ with a precision of \SI{10}{\kelvin}, \SI{0.05}{\dex}, and \num{0.015}\,R$_{\mathrm{J}}$ respectively. The p-T profile was well constrained, with the true profile less than \SI{20}{\kelvin} away from the posterior median. 
Contrary to the abundances of H$_{2}$O and CO which were well constrained with Gaussian posterior distributions, the abundances of FeH, K, and VO resulted in an upper bound, where the posterior maximum correctly identified an upper bound and lower values were not totally ruled out.

This can be attributed to the strong features produced by those molecules at short wavelengths, identified by the low-resolution SPHERE/IFS data. For the other molecules, we obtained correct upper bounds with the true abundances at lower values. 
The C/O ratio and metallicity were accurately retrieved within $1\sigma$ with a spread of \num{0.06} and \SI{0.09}{\dex} respectively.

Finally, when we applied \texttt{CROCODILE} to all three techniques (in green in Fig.~\ref{fig:doubleRetrieval_small_corner}), we obtained equivalent constraints on all parameters, with a negligible improvement on the spread of the C/O ratio and the accuracy of the metallicty.
Interestingly, the lower C/O ratio constrained by MRCCS relative to LRSP is similar to the inconsistent C/O ratios that have been reported for the HR~8799~b planet: \citet{Lavie2017ttHELIOSRETRIEVAL:/ttFormation} used photometric and low-resolution H- and K-band spectroscopic data from KECK/OSIRIS (binned down to a resolution of 60 to increase the S/N) and obtained a value close to unity, while \cite{Ruffio2021DeepSpectroscopy} derived a much lower value of 0.578 on medium-resolution ($R \approx 4000$) H and K band cross-correlation spectroscopic observations taken with the same instrument but processed differently. 

In summary, our test seemed to further confirm the systematics of the retrieval on K-band MRCCS data observed in the previous section, while these systematics disappear when combined with LRSP data.
In the next section, we investigate whether that is still the case when changing the input atmosphere.

\subsection{Exploration of the parameter space}
\label{sec:retrieval_study}

\begin{figure}[t]
    \centering
    \includegraphics[width=\hsize]{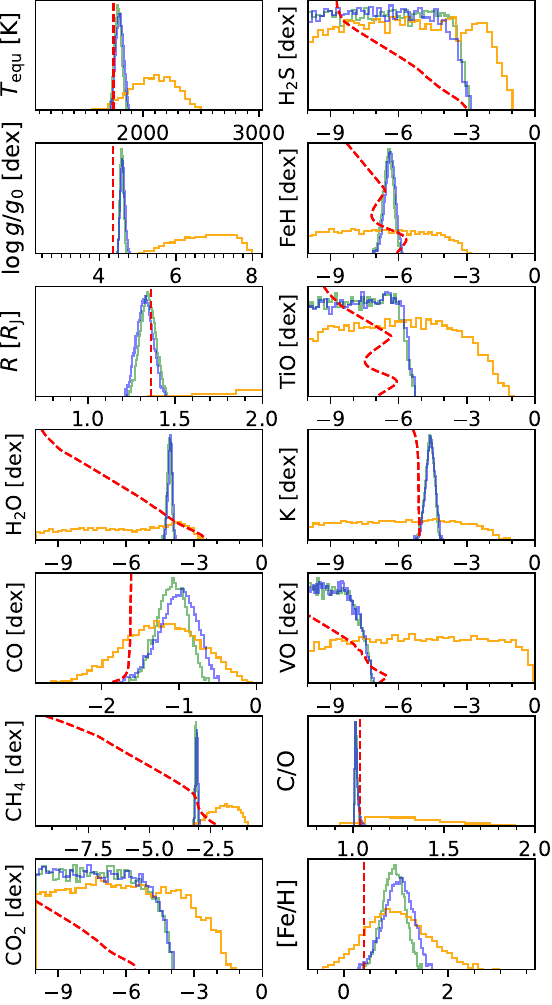}
    \caption{Posterior distributions obtained for the retrievals of our simulated atmosphere 5 (cf. Table~\ref{tab:simulated_values}) using \texttt{CROCODILE} on MRCCS data alone (orange), on LRSP data (blue), and on both MRCCS+LRSP combined (green). The input values are shown as dashed red lines.}
    \label{fig:summary_highlight_plot}
\end{figure}

\begin{figure*}[!ht]
    \centering
    \includegraphics[width=0.95\hsize]{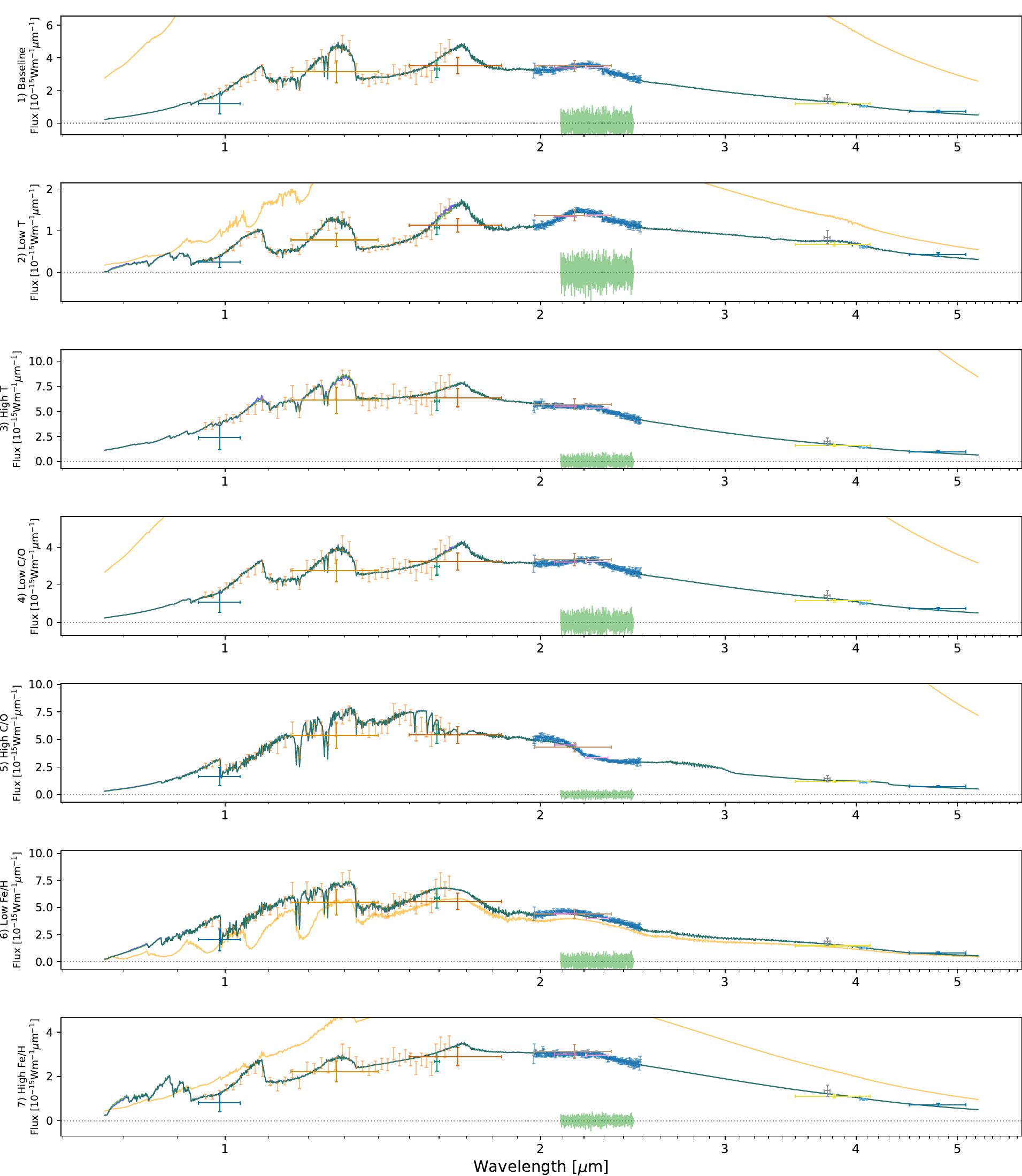}
    \caption{Comparison of the fitted spectra to our simulated datasets 1 to 7 (cf. Table~\ref{tab:simulated_values}) using \texttt{CROCODILE} on MRCCS data alone (orange), on LRSP data (blue), and on both MRCCS+LRSP combined (green). The fitted spectra were calculated using our forward model from the posterior median of the retrieved distributions shown in Fig.~\ref{fig:summary_plot}. The 11 colourful crosses represent the simulated photometric values and their corresponding uncertainty and equivalent width; the orange and blue error bars show the simulated $YH$-band VLT/SPHERE/IFS and K-band VLTI/GRAVITY spectra and corresponding uncertainties; and the contiuum-subtracted light green spectrum shows the simulated VLT/SINFONI MRCCS data.}
    \label{fig:summary_SED}
\end{figure*}

\begin{figure}[t]
    \centering
    \includegraphics[width=\hsize]{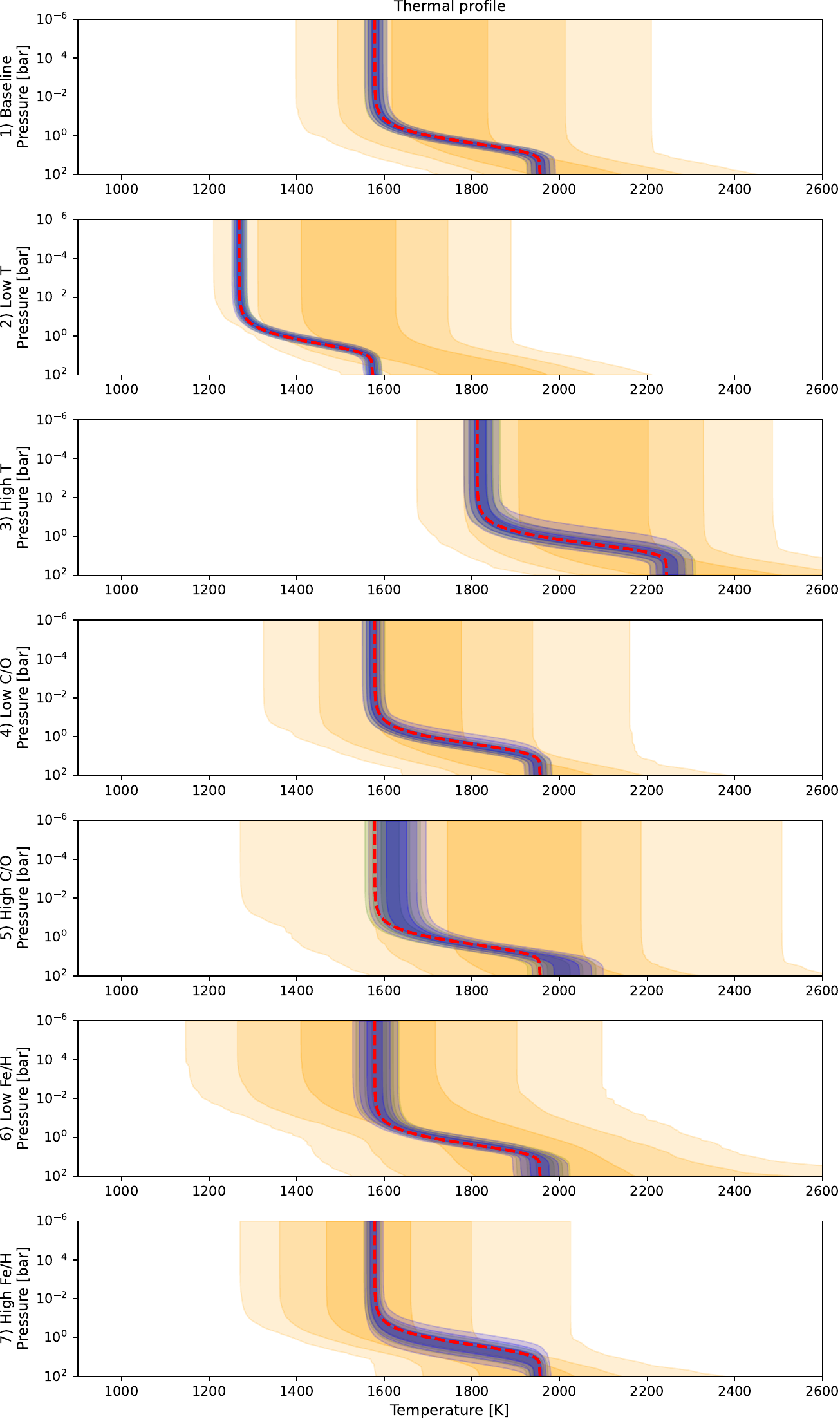}
    \caption{Posterior thermal profiles obtained for the retrievals of our simulated atmospheres 1 to 7 (cf. Table~\ref{tab:simulated_values}) using \texttt{CROCODILE} on MRCCS data alone (orange), on LRSP data (blue), and on both MRCCS+LRSP combined (green). The different envelopes of each colour represent the 
    68th, 95th, and 99th percentile credible intervals --- corresponding to $1\sigma$, $2\sigma$, and $3\sigma$ of a normal distribution --- of the posterior distributions of the retrieved thermal profiles. The input p-T profiles are shown as dashed red lines.}
    \label{fig:summary_PT}
\end{figure}

Since our results --- derived for a simple synthetic model of $\beta$~Pic~b --- might depend on the input atmosphere, we next explored how our framework fares in the parameter space around it as described in \ref{section:sim_target}. The model setup and priors used for our retrievals were the same as in the previous section (readers can refer to App.~\ref{app:retrieval_setup} for details). Fig.~\ref{fig:parameter_space_score} reports the scores of three metrics applied to our retrievals in order to compare the results of the three techniques. Fig~\ref{fig:summary_highlight_plot} focuses on the fifth atmospheric model as it is discussed below, while the fitted SEDs and posterior p-T profiles are shown in Fig.~\ref{fig:summary_SED} and Fig.~\ref{fig:summary_PT} respectively.  Figure~\ref{fig:summary_plot} in the appendix summarises all the resulting posterior distributions of the seven synthetic targets defined in Sect.~\ref{section:sim_target}, while the tables \ref{tab:retrieval_results_CC}, \ref{tab:retrieval_results_RES}, and \ref{tab:retrieval_results_CROCO} in App.~\ref{app:numerical_results} report the numerical values.


In order to quantitatively compare the results obtained by LRSP, MRCCS, or LRSP+MRCCS, we computed a slightly modified Mahalanobis distance $d_{\theta}$ (Eq.~\ref{equ:mahalanobis}), the relative error $\epsilon_{\theta}$ (Eq.~\ref{equ:rel_err}), and the mean squared error $\mathrm{MSE}_{\theta}$ (Eq.~\ref{equ:mse}) from the marginalised posterior distribution of each parameter $\theta$ and for each simulated atmosphere:

\begin{align}
    d_{\theta} &= (\hat{\theta} - \theta_{\mathrm{T}})/\sigma_{\theta} \label{equ:mahalanobis}\\
    \epsilon_{\theta} &= (\hat{\theta} - \theta_{\mathrm{T}})/\theta_{\mathrm{T}} \label{equ:rel_err}\\
    \mathrm{MSE}_{\theta} &= \frac{1}{N_{\mathrm{S}}}\sum_{i=1}^{N_{\mathrm{S}}}(\theta_{i}-\theta_{\mathrm{T}})^{2} \label{equ:mse} \,.
\end{align}

In the equations above, $\hat{\theta}$ is the posterior median, $\theta_{\mathrm{T}}$ is the ground truth of parameter $\theta$, $\sigma_{\theta}$ is the distance of the median from the closest bound of the 68th percentile confidence interval (i.e. if $\theta < \hat{\theta}$, then $\sigma_{\theta}$ is equal to the difference between the median and the 16th percentile, and the difference between the median and the 84th percentile otherwise), and $\theta_{i}$ denotes the $N_{\mathrm{S}}$ points sampled by the nested algorithm. We choose these three metrics for the following reasons. First, the Mahalanobis metric measures the distance from the ground truth in terms of standard deviations (more precisely here in terms of 68th percentile confidence interval), meaning that a bad score is obtained when the uncertainty is underestimated. Second, the relative error measures the accuracy with which the parameters are retrieved. Finally, the mean squared error measures the quality of the median posterior as an estimator for the ground truth as it incorporates both its variance and its bias. Since the true molecular abundances have non-constant vertical profiles, $\theta_{\mathrm{T}}$ represents the point of the true profile closest to the retrieved value, that is $(\hat{\theta} - \theta_{\mathrm{T}})$ is zero if there is an atmospheric layer where the retrieved abundance matches the true value. Figure~\ref{fig:parameter_space_score} summarises the results of these three metrics for each retrieval case and for each technique (MRCCS, LRSP, or MRCCS+LRSP, i.e. \texttt{CROCODILE}), and by averaging over three groups of parameters: the molecular abundances, the physical parameters $T_{\mathrm{equ}}$, $\log(g)$, and $R$, and the C/O ratio and metallicity.

The retrieval using MRCCS data alone resulted in the worst overall score with respect to every metric and for every subgroup of parameters, except for the molecular abundances with the Mahalanobis metric. This can be explained by the broad posterior distributions obtained with the technique, meaning that the ground truth was more often contained within the 68th percentile confidence interval although the retrieved value was significantly far from the true value, especially for $T_{\mathrm{equ}}$, $\log(g)$, and $R$. Combining MRCCS with LRSP data lead to the lowest relative error and mean squared error for all groups of parameters, and yielded approximately equal Mahalanobis distance than the retrievals on LRSP-only data. Overall, that is across all the cases, the combined fit on LRSP and MRCCS data achieved to retrieve 68\% of the parameters within $0.7\sigma$, against $0.6\sigma$ for the LRSP-only fit and $1.1\sigma$ for the MRCCS-only fit, while the same proportion of the parameters was retrieved with a relative error of less than $6\%$ for LRSP+MRCCS, against $8\%$ for LRSP and $28\%$ for MRCCS. Hence, the combined fit on MRCCS and LRSP data was overall slightly more accurate and resulted in less underestimation of the uncertainty than the fit on LRSP data only, while the fit on MRCCS data only was largely inaccurate.

\begin{figure}[h]
    \centering
    \includegraphics[width=\hsize]{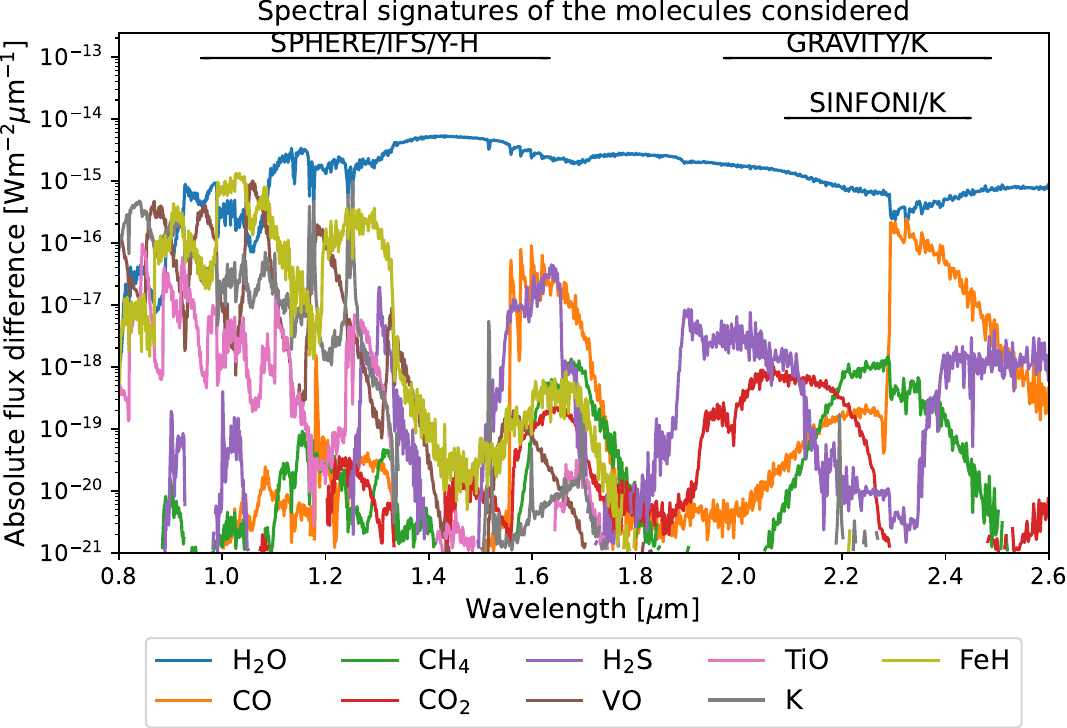}
    \caption{Absolute difference between the spectrum calculated with the full chemical model used for our simulated atmosphere 1 (as described in Table~\ref{tab:simulated_values}) and spectra where each molecule included in our framework was removed one after the other. For example water shows a large difference due to its absorption features throughout the whole range. The black lines indicate the spectral coverage of the different spectroscopic instruments included in our tests.}
    \label{fig:molecular_signatures}
\end{figure}

It is worth noting which molecule could be constrained and in what scenario. While the abundances of H$_{2}$O and CO resulted in narrow posterior distributions, those of CO$_{2}$, H$_{2}$S, TiO, and K almost were never retrieved accurately, with at most upper bounds and only a good constraint of K at high C/O ratio. This can be explained by the strong absorption features of H$_{2}$O and CO included in the spectral range covered by our synthetic data, while this is not the case for CO$_{2}$, H$_{2}$S, TiO, and K. This can be seen in Fig.~\ref{fig:molecular_signatures}, where we illustrate the spectral signatures of the molecules included in our retrieval by computing the absolute difference in flux between the full chemical model at equilibrium and when each molecule is removed successively. The abundance of CH$_{4}$ seemed to be retrievable when it was significantly higher in the input atmosphere, namely at low temperature and at high C/O ratio. Finally, FeH, and VO obtained similar results with mostly correct upper bounds and a few good constraints of FeH at high C/O ratio and low metallicity when LRSP data was included, which can be explained by the spectral features of FeH and VO that are covered by the VLT/SPHERE/IFS $Y-H$-band LRS data but were not included in the MRCCS data.

Looking at Fig.~\ref{fig:parameter_space_score} case by case, it seems that there is one case that yielded relatively worst scores than the others when including LRSP data (i.e. LRSP alone or combined with MRCCS), namely at high C/O ratio (case 5). Indeed, all three subgroups of parameters resulted in relatively higher Mahalanobis distance and relative error than the other cases. This can be understood better in Fig.~\ref{fig:summary_plot}: the retrieved surface gravity, the abundances of CO and K as well as the metallicity all overestimated the ground truth by a few $\sigma$. However, this is more due to the small spread of the posterior distributions (as is the case for the surface gravity), quickly resulting in large Mahalanobis distance, or to the fact that the ground truth itself is a small number (for example the metallicity), also resulting in a large relative error. Looking at the mean squared error, the fifth case actually obtained fairly similar scores to the others.


Finally, the quality of an atmospheric retrieval is usually estimated from the fit to the input spectrum, which we provide in Fig.~\ref{fig:summary_SED}. Remembering that all our datasets contained random scatter corresponding to their errorbar, the spectral fit --- computed using the median of the posterior distributions of each parameter --- showed no significant deviation from the LRSP data when it was included in the fit. Non-surprisingly, the retrieval using only MRCCS data did not match the LRSP data due to its inability to constrain the continuum and thus the thermal profile of the atmosphere. Indeed, as shown in Fig.~\ref{fig:summary_PT}, the temperature-pressure profile was largely unconstrained by the MRCCS data with a $1\sigma$ envelope of \SI{200}{\kelvin} mostly overestimating the input profile except in the cases 6 and 7. Both the fit on LRSP data only and the combined fit were able to precisely constrain the thermal profile.

\section{Discussion}
\label{sec:discussion}

\subsection{Applicability of \texttt{CROCODILE}}
\label{sec:applicability}

\begin{figure}[t]
    \centering
    \includegraphics[width=\hsize]{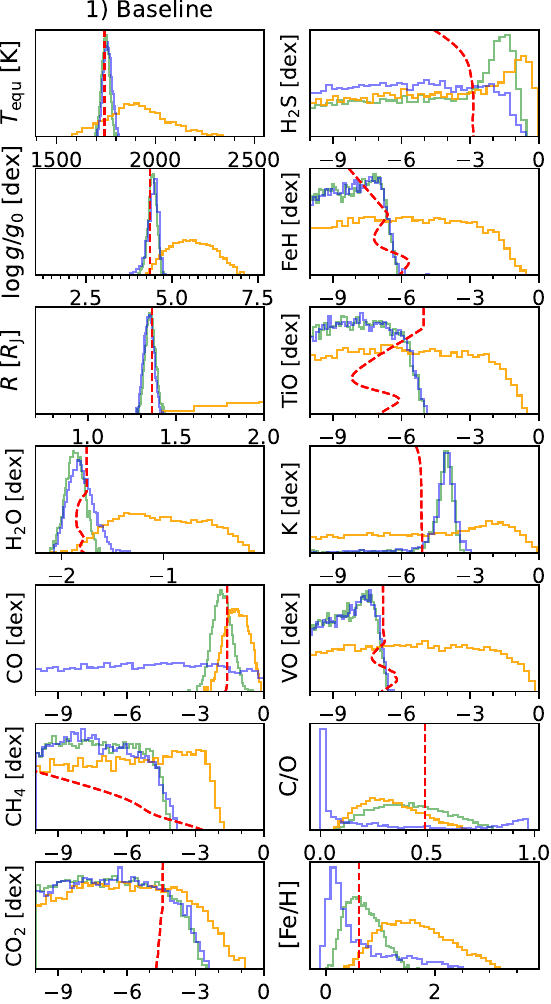}
    \caption{Posterior distributions obtained when removing the synthetic VLTI/GRAVITY data from the atmospheric fit. The input atmosphere is the same as case 1 (cf. Table~\ref{tab:simulated_values}), and as in Fig.~\ref{fig:summary_highlight_plot} we also compare the results when applying \texttt{CROCODILE} to MRCCS data alone (orange), LRSP data (blue), and on both MRCCS+LRSP combined (green). However, this time the low-resolution spectroscopy only included synthetic VLT/SPHERE data in the $Y-H$ bands. The input values are shown as dashed red lines.}
    \label{fig:posterior_no_GRAVITY}
\end{figure}

The results presented in the last section show the robustness of the framework of \texttt{CROCODILE} to retrieve the parameters of a diverse population of gas giant atmospheres across a range of temperatures and chemical composition. However, our analysis also showed that the addition of K-band SINFONI MRCCS data along the list of LRSP data in the retrieval did not significantly improve our results: the posterior distributions mostly matched, with a slight gain of accuracy and better estimation of the uncertainty as reported in Sect.~\ref{sec:retrieval_study}. We argue that this is because of the high-quality K-band VLTI/GRAVITY spectrum at a spectral resolution of $R\approx500$ which can already constrain the abundance of CO on its own, and hence the medium-resolution spectrum provided by VLT/SINFONI does not add much information that is not already there. To demonstrate this point, we repeated the retrievals on case 1, however this time we excluded the GRAVITY spectrum from the fit. The results of this final test are presented in Fig.~\ref{fig:posterior_no_GRAVITY} and in Table~\ref{tab:retrieval_results_no_GRAVITY} in App.~\ref{app:numerical_results}. Indeed, the low-resolution Y- to H-band VLT/SPHERE spectrum together with the photometric points were not enough on their own to constrain the molecular abundance of CO, leading to a mostly undetermined C/O ratio and a poorly constrained metallicity. When the K-band SINFONI data was added to the fit, the difference was significant: the abundance of CO became well constrained, leading to a good estimation of the C/O ratio (C/O$=0.41^{+0.17}_{-0.16}$ against the input value of 0.49) and accurate constraint of the metallicity (Fe/H$=0.61^{+0.30}_{-0.23}$ against the input value of 0.61).

While Sect.~\ref{sec:retrieval_study} demonstrated the robustness of \texttt{CROCODILE} against different input atmospheres, this last test showed its actual strength. Namely, it allows the combination of low signal-to-noise cross-correlation spectroscopic data together with a few photometric points and low-resolution spectroscopic data to measure the molecular abundances of closely separated directly imaged low-mass companions, without the need for high-quality spectroscopic data such as measured by VLTI/GRAVITY.

\subsection{Future opportunities}

The major disadvantage of removing the continuum to perform cross-correlation with spectral templates is made obvious in Fig.~\ref{fig:summary_plot}: the information carrying the temperature is mostly lost with the continuum, rendering its determination almost impossible using only cross-correlation and further degrading the determination of the other parameters such as surface gravity, C/O ratio, and metallicity. It therefore seems that the continuum should not be removed. If it is true that molecular lines present in substellar companions differ enough from the atomic lines found in stellar spectra, then this fact stays true if the continuum of the planet can be retrieved. Several studies are indeed able to retrieve the continuum of directly imaged low-mass companions using medium-resolution spectroscopy: for example \cite{Daemgen2017HighHD106906b} and \cite{Petrus2021Medium-resolutionB} were able to measure the spectrum of the planetary-mass objects HD 106906 b and HIP 65426 b using VLT/SINFONI, while most recently \cite{Miles2022TheB} released the spectrum of the young brown dwarf VHS 1256 b taken with the NIRSpec IFU and MIRI MRS onboard \texttt{JWST}, the highest fidelity spectrum to date of a planetary-mass object, which will undoubtedly yield the most robust derivation of the atmospheric properties of a planetary-mass object to date.

The point of our study is not that the continuum of those spectra could have been removed with similar constraints for the atmosphere of the exoplanet, but rather to enable the study of close-in low-mass companions where the continuum cannot be extracted at all (or at least not within reasonable integration times) due to contamination by stellar speckles, as is the case in \cite{Hoeijmakers2018Medium-resolutionImaging}, \cite{Cugno2021MolecularSystem}, \cite{Ruffio2021DeepSpectroscopy}, and \cite{Wang2021DetectionSpectroscopy}. Indeed, future direct-imaging searches for exoplanets will focus more and more on the innermost regions of planetary systems, where radial velocity searches indicate that the occurrence rate of planets is higher \citep{Cumming2008ThePlanets,Fulton2021CaliforniaLine}. However, at such close separations, the location of a planet is only at a few resolution elements from the central star. There, the stellar PSF is still so intense that extracting medium and high spectral resolution spectra for such planets will come at the cost of the continuum --- at least initially. That is when \texttt{CROCODILE} --- and similar statistical frameworks based on the cross-correlation of continuum-removed spectra with spectral models~\citep[e.g.][]{Ruffio2021DeepSpectroscopy} --- comes into play by enabling a full Bayesian analysis to characterise more and more challenging directly imaged exoplanets.

\section{Summary and conclusion}
\label{sec:conclusion}

The results of this work are summarised in the following:

\begin{itemize}
    \item We adapted the statistical framework of \citet{Brogi2019RetrievingSpectroscopy} from exoplanet transit spectroscopy to the different observational techniques of directly imaged exoplanets, including photometry, low-resolution spectroscopy, and medium (and higher) resolution cross-correlation spectroscopy, which can constrain the atmospheric properties of gas giant exoplanets such as temperature-pressure profile and chemical composition. Our Python routine, called \texttt{CROCODILE}, is freely available for download on GitHub.\footnote{\url{https://github.com/JHayoz/CROCODILE}}
    \item We thoroughly tested our framework, first on a synthetic atmosphere created with the same input as our forward model to verify the framework, and then on a grid of atmospheres at chemical equilibrium to test the ability of our free forward model to robustly constrain atmospheric properties in a realistic scenario.
    \item We find that using medium-resolution cross-correlation spectroscopy alone leads to a generally poor constraint of the atmospheric parameters due to the loss of the continuum. The results improved a lot when fixing the planetary radius to its ground truth. However, we noticed strong correlations between the radius and the surface gravity and the molecular abundances of H$_{2}$O and CO This means that fixing the radius to a wrong value might strongly bias the retrieved values.
    \item We further find that combining medium-resolution cross-correlation spectroscopy to photometry and low-resolution spectroscopy leads to an accurate determination of the atmospheric parameters. Indeed, 68\% of the parameters were retrieved within $0.9\sigma$ of the ground truth, with no systematical bias despite the different chemical model used in the forward model and in the simulated data.
    \item Our last test shows that low-quality (i.e. high noise) medium-resolution ($R \approx 4000$) K-band cross-correlation spectroscopy, when combined with photometry, leads to an equivalently accurate constraint of the abundance of CO than high-quality (i.e. low noise) low-resolution ($R \approx 500$) K-band spectroscopy. To put into perspective, the actual observations that our synthetic data is based on was acquired in 2.5 hrs of integration time for VLT/SINFONI on one telescope \citep{Hoeijmakers2018Medium-resolutionImaging} versus 1.4 hrs for VLTI/GRAVITY on four telescopes \citep{Nowak2020PeeringInterferometry}, that is the medium-resolution cross-correlation spectroscopy required less than half of the total telescope time that was required for the low-resolution spectro-interferometry.
    \item Our work shows that medium-resolution cross-correlation spectroscopy in the K band is best interpreted alongside $Y-M$-band photometry and, if available, low-resolution spectroscopy in the $Y-H$ bands, to robustly infer the atmospheric properties of gas giant exoplanets.
    \item In summary, \texttt{CROCODILE} provides the statistical framework to interpret medium- (and possibly higher-) resolution spectroscopic data alongside photometric and low-resolution spectroscopic data of directly imaged gas giant exoplanets at close separations to their host star, where the continuum of their spectrum can only be extracted reliably at low spectral resolution due to the pollution by the stellar PSF. \texttt{CROCODILE} allows the measurement of the atmospheric properties of gas giants such as the pressure-temperature profile and the abundances of individual molecules. In the future, we plan to add the option to choose between different thermal, chemical, and cloud models, in order to allow model selection via computation of Bayes' factor between competing models.
\end{itemize}

With the arrival of the next generation of exoplanet direct-imaging instruments such as ERIS at the VLT, MIRI onboard \texttt{JWST}, and the upcoming METIS at the ELT, it is crucial to define robust statistical frameworks that take into account the intricacies of the different observational and post-processing techniques, while acknowledging the risk of model degeneracies. In particular, such frameworks should be validated on a representative subset of the parameter space of exoplanetary atmospheres to identify the limits of applicability of the forward model before being carried out on real data.

\begin{acknowledgements}
      The authors thank the anonymous referee for their constructive insights which helped to improve the quality of this work. JH and SPQ gratefully acknowledge the financial support from the Swiss National Science Foundation (SNSF) under project grant number 200020\_200399. GC, PP, EOG thank the SNSF for financial support under grant numbers P500PT\_206785 and 200020\_200399. MJB acknowledges the financial support from ETH Zurich. This work has been carried out within the framework of the NCCR PlanetS supported by the Swiss National Science Foundation under grants 51NF40\_182901 and 51NF40\_205606.
      Some of our plots were made using the Python package \texttt{corner.py} developed by \cite{Foreman-Mackey2016Corner.py:Python}. 
      \textbf{Contributions:} the authors listed in this article are ordered according to the following: JH built the framework, did the analysis, and wrote the paper; GC, SPQ, and PP provided feedback and supervision during the whole project; while the rest of the co-authors, who are ordered alphabetically, provided feedback in later stages of the project.
\end{acknowledgements}

\FloatBarrier

\bibliographystyle{aa}
\bibliography{references-fixed}

\begin{appendix}

\section{Derivation of the log-likelihood function for cross-correlation spectroscopy}
\label{app:derivation_CCS}

In the following, we provide the derivation of the log-likelihood function $\log \Lagr_{\mathrm{CCS}}$ given in Eq.~\ref{equ:likelihood_CCS} originally developed by \citet{Brogi2019RetrievingSpectroscopy}, but adapted to our notation. Let $d=\left(d_{1},...,d_{N}\right)$ be the measured continuum-removed spectrum of an observed exoplanet with $N$ spectral channels and $m\left(\theta,v_{\mathrm{R}}\right) =: m = \left(m_{1},...,m_{N}\right)$ be the associated forward model. For the sake of readability, we drop the dependency on the model parameters $\theta$ and the radial velocity $V_{\mathrm{R}}$ in the following. We assume that the data is of the form $d=a m + n$, with a scaling factor $a$ and noise $n=\left(n_{1},...,n_{N}\right)$, where the components obey a normal distribution with a common standard deviation $\sigma$, that is $n_{i} \sim N\left(0,\sigma\right)$. \citet{Brogi2019RetrievingSpectroscopy} choose $a=1$ and scale their model spectrum by the stellar flux and the area ratio between the planet and the star. In our study, the data and the forward model were created with the same scaling factor, therefore $a=1$ also holds here. However, if the observed spectrum is provided in arbitrary units, then one could replace the multiplicative factor $R^{2}/d^{2}$ used in our forward model by $a$ and treat is as a nuisance parameter.

The likelihood function associated to the noise statistics is given by 

\begin{equation}
    \Lagr = \left(\frac{1}{\sqrt{2\pi\sigma^2}}\right)^{N} \exp\left(-\sum_{i=1}^{N}\frac{\left(d_{i}-m_{i}\right)^{2}}{2\sigma^{2}}\right)\,,
\end{equation}

which leads to the following log-likelihood function (ignoring constant additive terms):

\begin{equation}
\label{equ:lagrange_app}
    \log \Lagr \approx - N \log \sigma - \frac{1}{2\sigma^{2}}\sum_{i=1}^{N}\left(d_{i}-m_{i}\right)^{2}\,.
\end{equation}

The variable $\sigma$ represents the uncertainty associated to each spectral channel and is assumed to be constant for all spectral channels. It is challenging to compute due to the pollution by telluric absorption lines or stellar light. Therefore, we replace it by its maximum likelihood estimator $\hat{\sigma}$:

\begin{align}
    0 &= \frac{\partial \log\Lagr}{\partial \sigma} = -\frac{N}{\sigma} + \frac{1}{\sigma^{3}}\sum_{i=1}^{N}\left(d_{i}-m_{i}\right)^{2}\notag\\
    \hat{\sigma}^{2} &= \frac{1}{N}\sum_{i=1}^{N}\left(d_{i}-m_{i}\right)^{2}\,.
\end{align}

By inserting $\hat{\sigma}$ back into Eq.~\ref{equ:lagrange_app} and expanding the product $\left(d_{i}-m_{i}\right)^{2} = d_{i}^{2} - 2d_{i}m_{i} + m_{i}^{2}$, we finally obtain:

\begin{align}
    \log \Lagr &\approx -\frac{N}{2}\log\left(\frac{1}{N}\sum_{i=1}^{N}\left(d_{i}^{2} - 2d_{i}m_{i} + m_{i}^{2}\right)\right) - \frac{N}{2}\notag\\
     &\approx -\frac{N}{2}\log \left( s_{d}^{2} - 2K + s_{m}^{2}\right)\,,
\end{align}

with the definitions

\begin{align}
    s_{d}^{2} &= \frac{1}{N}\sum_{i=1}^{N}d_{i}^{2} \label{equ:sd2} \\
    s_{m}^{2} &= \frac{1}{N}\sum_{i=1}^{N}m_{i}^{2} \label{equ:sm2} \\
    K &= \frac{1}{N} \sum_{i=1}^{N} d_{i}m_{i}\,,
\end{align}

which corresponds to Eq.~\ref{equ:likelihood_CCS}. For the sake of clarity, we note here again that instead of the cross-covariance $K$, one can use the normalised cross-correlation $C$ by using Eq.~\ref{equ:K_C} and inserting it into the log-likelihood function:

\begin{align}
    \log \Lagr &\approx -\frac{N}{2}\log \left( s_{d}^{2} - 2K + s_{m}^{2}\right) \\
    &\approx -\frac{N}{2}\log \left( s_{d}^{2} - 2C\sqrt{s_{d}^{2}s_{m}^{2}} + s_{m}^{2}\right) \,.
\end{align}

\section{The retrieval setup}
\label{app:retrieval_setup}

Our atmospheric retrieval was setup in the following configuration. We set the equilibrium temperature $T_{\mathrm{equ}}$ and logarithm of the surface gravity $\log g$ as free retrievable parameters governing the thermal structure. Implicitly, the thermal structure from \citet{Guillot2010OnAtmospheres} contains four more parameters which were kept fixed in the simulated datasets and in the forward model: the average opacity in the infrared $\kappa_{\mathrm{IR}}$ = \SI{0.01}{\centi\meter\squared\per g}, the ratio of the average opacity in the optical and infrared $\gamma=\kappa_{\mathrm{V}}/\kappa_{\mathrm{IR}} = 0.4$, the interior temperature $T_{\mathrm{int}}$ = \SI{200}{\kelvin}, and the bottom pressure $P_{0}$ = \SI{100}{\bar}. For the free parameters governing the chemical composition, we included the logarithms of the mass fractions (assumed to be vertically constant) of the following molecules: H$_{2}$O, CO, CO$_{2}$, CH$_{4}$, H$_{2}$S, FeH, TiO, K, and VO. Finally, we set the abundances of H$_{2}$ and He equal to \SI{75}{\percent} and \SI{25}{\percent} of the remaining mass fraction in each layer. The last free parameter of the model was the radius of the planet $R$. The priors used in this study are given in Table~\ref{tab:prior}.

\begin{table}[htb!]
    \centering
    \caption{Priors used.}
    \renewcommand*{\arraystretch}{1.7}
    \begin{tabular}{l l}
    \toprule
    Parameter & Prior \\
    \midrule
    T$_{\mathrm{equ}}$ & $\mathcal{U}(0,5000)$ \\
    $\log(g/g_{0})$    & $\mathcal{U}(1,8)$    \\
    $R$                & $\mathcal{U}(0.1,10)$ \\
    $\log$(H$_{2}$O)   & $\mathcal{U}(-10,0)$  \\
    $\log$(CO)         & $\mathcal{U}(-10,0)$  \\
    $\log$(CH$_{4}$)   & $\mathcal{U}(-10,0)$  \\
	\bottomrule
    \end{tabular}
    \hspace{4mm}
    \begin{tabular}{l l}
    \toprule
    Parameter & Prior \\
    \midrule
    $\log$(CO$_{2}$)       & $\mathcal{U}(-10,0)$  \\
    $\log$(H$_{2}$S)   & $\mathcal{U}(-10,0)$  \\
    $\log$(FeH)        & $\mathcal{U}(-10,0)$  \\
    $\log$(TiO)        & $\mathcal{U}(-10,0)$  \\
    $\log$(K)          & $\mathcal{U}(-10,0)$  \\
    $\log$(VO)         & $\mathcal{U}(-10,0)$  \\
	\bottomrule
    \end{tabular}
    \label{tab:prior}
\end{table}
\FloatBarrier

\section{Derivation of the optimal size of the Gaussian filter}
\label{app:filter_size}

\begin{figure}[t]
    \centering
    \includegraphics[width=\hsize]{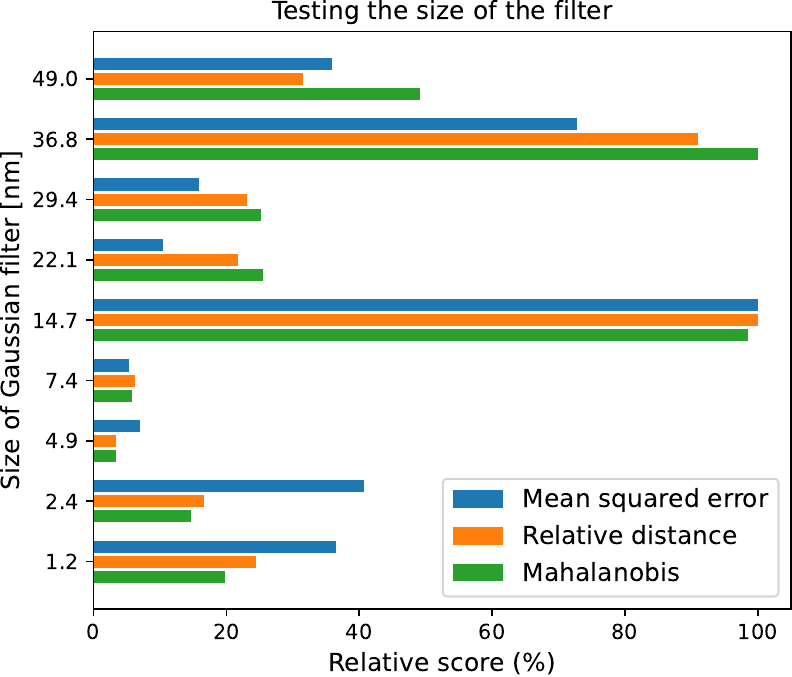}
    \caption{Scores for different sizes of the Gaussian filter used to remove the continuum of the synthetic medium-resolution spectroscopic data used in this work.}
    \label{fig:score_filter_size}
\end{figure}

We derived the optimal size of the Gaussian filter used to remove the continuum of our synthetic medium-resolution spectroscopic data with the following test. We repeated the setup of Sect.~\ref{sec:preliminary_test} on MRCCS data alone for filter sizes of 1.2, 2.4, 4.9, 7.4, 14.7, 22.1, 29.4, 36.8, and 49 nm (corresponding to 5, 10, 20, 30, 60, 90, 120, 150, and 200 spectral bins at the spectral resolution of VLT/SINFONI), and computed the same three metrics as in Sect.~\ref{sec:retrieval_study}. To obtain an overall score for each of the three metrics, we multiplied the results for each parameter together, and divided by the worst score to give the results as percentages. The result of this analysis is shown in Fig.~\ref{fig:score_filter_size}. The filter size at 4.9 nm outperformed all other filters with respect to the Mahalanobis and relative distance, and was a close second in terms of mean squared error (7.0\% against 5.4\%).

\section{Results: Verification of the framework}
\label{app:preliminary_test_plot}
\begin{figure}[t]
    \centering
    \includegraphics[width=\hsize]{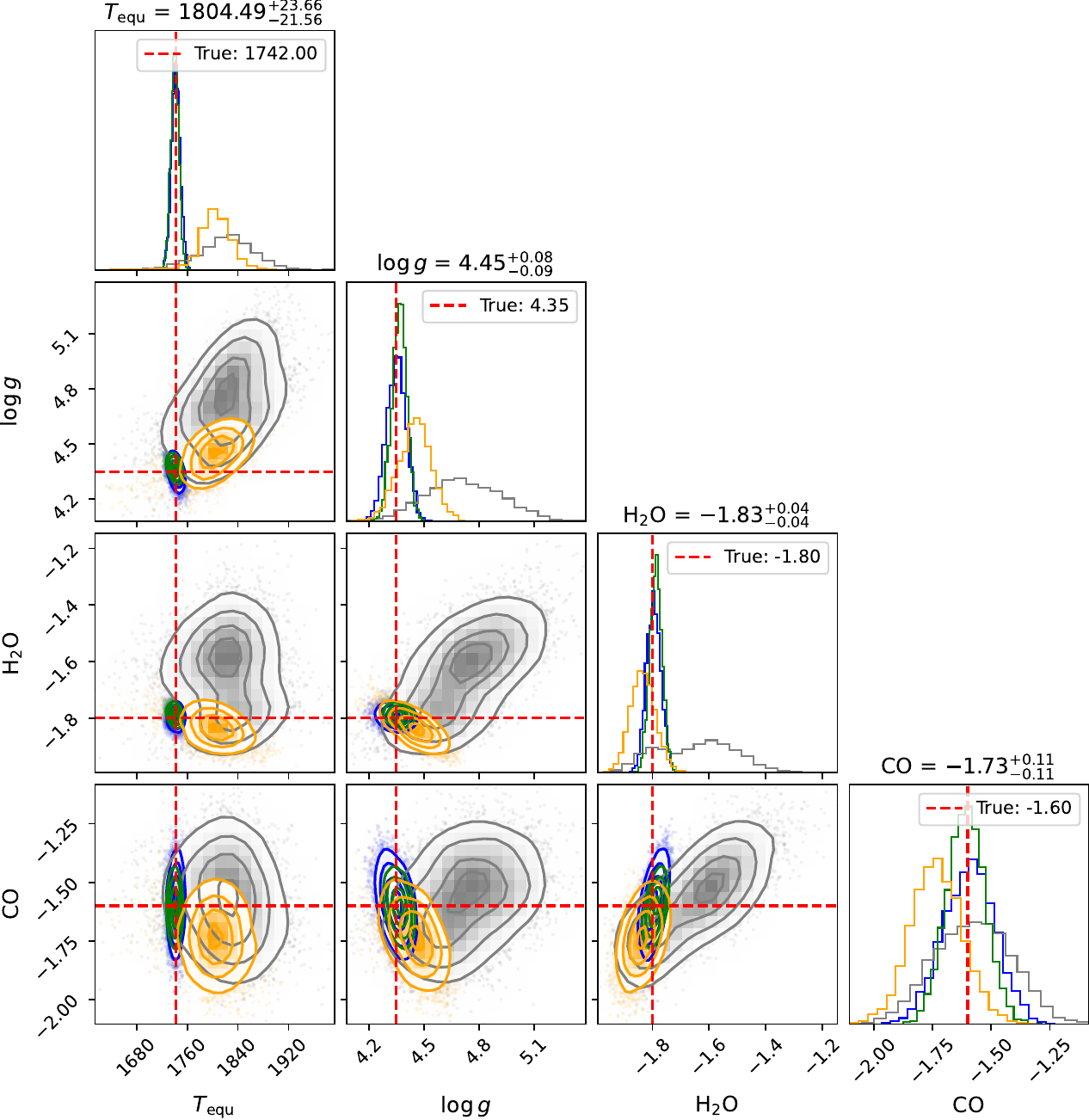}
    \caption{Resulting posterior distributions of our preliminary test on MRCCS data when fixing the radius to its simulated value (orange), compared to the previous results with the radius included as free parameter (grey: old results on MRCCS data, blue: only LRSP, green: MRCCS and LRSP). The dashed red line shows the ground truth used as input.}
    \label{fig:preliminary_test_techniques_noR}
\end{figure}

\FloatBarrier

\section{Results: Exploration of the parameter space}
\label{app:param_space_plots}

\begin{figure*}[t]
    \centering
    \includegraphics[width=\hsize]{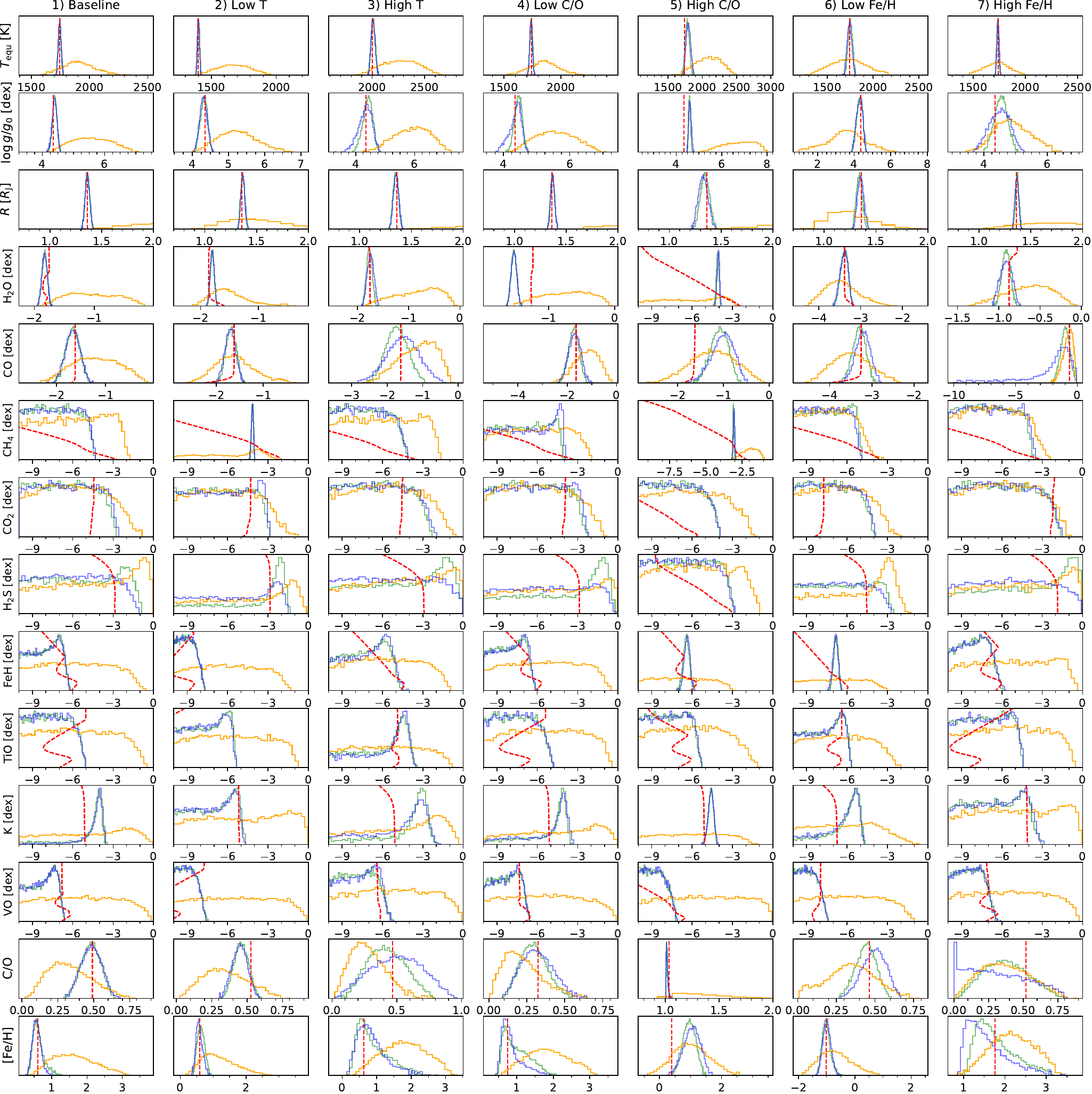}
    \caption{Comparison of the posterior distributions obtained for the retrievals of our simulated atmospheres 1 to 7 (cf. Table~\ref{tab:simulated_values}) using \texttt{CROCODILE} on MRCCS data alone (orange), on LRSP data (blue), and on both MRCCS+LRSP combined (green). The posterior distributions are shown as histograms of the sampled parameters and are to be read row by row for each model parameter and column by column to compare to the other input atmospheres. The input values are shown as dashed red lines. The input molecular abundances --- which were calculated at chemical equilibrium and therefore vary vertically with the pressure and the temperature --- are shown as functions of the atmospheric layers between $10^{2}$ and \SI{1e-6}{\bar} in logarithmic scale. The posterior distributions of the C/O ratio and metallicity Fe/H are inferred from the posterior distributions of the molecular abundances.}
    \label{fig:summary_plot}
\end{figure*}
\FloatBarrier
%
%

\section{Detailed numerical results of our study}
\label{app:numerical_results}

\begin{table*}[htb!]
    \centering
    \renewcommand*{\arraystretch}{1.7}
    \caption{List of posterior median and 68th percentile credible interval for our verification of the statistical framework of \texttt{CROCODILE} (cf. Sect.~\ref{sec:preliminary_test}).}
    \begin{tabular}{l l|c c c c c}
    \toprule
    Parameter & & Truth & MRCCS & LRSP & MRCCS+LRSP & MRCCS w/o R \\
    \midrule
    T$_{\mathrm{equ}}$ & [K] & 1742 & 1823$^{+41}_{-46}$ & 1742$^{+7}_{-7}$ & 1740$^{+6}_{-6}$ & 1804$^{+24}_{-22}$ \\
    $\log(g/g_{0})$ & [dex]  & 4.35 & 4.71$^{+0.22}_{-0.21}$ & 4.35$^{+0.05}_{-0.05}$ & 4.37$^{+0.04}_{-0.04}$ & 4.45$^{+0.08}_{-0.09}$ \\
    $R$ & [R$_{\mathrm{J}}$] & 1.36 & 1.56$^{+0.18}_{-0.15}$ & 1.36$^{+0.01}_{-0.01}$ & 1.36$^{+0.01}_{-0.01}$ & 0.00$^{+0.00}_{-0.00}$ \\
    H$_{2}$O & [dex]         & -1.8 & -1.64$^{+0.13}_{-0.17}$ & -1.80$^{+0.03}_{-0.03}$ & -1.79$^{+0.02}_{-0.02}$ & -1.83$^{+0.05}_{-0.04}$ \\
    CO & [dex]               & -1.6 & -1.57$^{+0.17}_{-0.17}$ & -1.59$^{+0.11}_{-0.11}$ & -1.61$^{+0.08}_{-0.08}$ & -1.73$^{+0.11}_{-0.11}$ \\
    \bottomrule
    \end{tabular}
    \label{tab:retrieval_results_preliminary}
\end{table*}

\begin{table*}[htb!]
    \centering
    \caption{List of posterior median and 68th percentile credible interval for the retrievals based only on MRCCS data.}
    \renewcommand*{\arraystretch}{1.7}
    \begin{tabular}{l l|c c c c c c c}
    \toprule
    Parameter & & 1) Baseline & 2) Low T & 3) High T & 4) Low C/O & 5) High C/O & 6) Low Fe/H & 7) High Fe/H \\
    \midrule
    T$_{\mathrm{equ}}$ & [K] & 1900$^{+126}_{-115}$ & 1675$^{+120}_{-119}$ & 2272$^{+159}_{-167}$ & 1850$^{+110}_{-111}$ & 2102$^{+160}_{-177}$ & 1727$^{+168}_{-171}$ & 1732$^{+100}_{-112}$ \\
    $\log(g/g_{0})$ & [dex]  & 5.55$^{+0.62}_{-0.58}$ & 5.20$^{+0.46}_{-0.43}$ & 5.98$^{+0.49}_{-0.58}$ & 5.53$^{+0.57}_{-0.52}$ & 6.85$^{+0.63}_{-0.81}$ & 3.59$^{+0.73}_{-0.77}$ & 4.83$^{+0.51}_{-0.41}$ \\
    $R$ & [R$_{\mathrm{J}}$] & 2.84$^{+2.10}_{-0.87}$ & 1.49$^{+0.41}_{-0.25}$ & 4.34$^{+1.91}_{-1.86}$ & 3.24$^{+1.63}_{-1.07}$ & 4.30$^{+2.76}_{-1.76}$ & 1.21$^{+0.20}_{-0.14}$ & 1.90$^{+0.59}_{-0.45}$ \\
    H$_{2}$O & [dex]         & -1.07$^{+0.49}_{-0.38}$ & -1.58$^{+0.28}_{-0.21}$ & -0.71$^{+0.32}_{-0.52}$ & -0.69$^{+0.32}_{-0.40}$ & -5.84$^{+2.13}_{-2.59}$ & -3.50$^{+0.33}_{-0.29}$ & -0.60$^{+0.22}_{-0.28}$ \\
    CO & [dex]               & -1.21$^{+0.45}_{-0.42}$ & -1.64$^{+0.32}_{-0.31}$ & -1.02$^{+0.38}_{-0.53}$ & -1.13$^{+0.37}_{-0.44}$ & -1.22$^{+0.40}_{-0.41}$ & -3.51$^{+0.37}_{-0.36}$ & -0.67$^{+0.27}_{-0.35}$ \\
    CH$_{4}$ & [dex]         & -5.74$^{+2.44}_{-2.83}$ & -5.02$^{+1.49}_{-3.37}$ & -5.96$^{+2.77}_{-2.66}$ & -6.06$^{+2.34}_{-2.49}$ & -1.89$^{+0.48}_{-0.58}$ & -7.41$^{+1.66}_{-1.69}$ & -6.73$^{+2.07}_{-2.04}$ \\
    CO$_{2}$ & [dex]         & -6.17$^{+2.48}_{-2.33}$ & -6.31$^{+2.38}_{-2.35}$ & -5.90$^{+2.55}_{-2.58}$ & -6.17$^{+2.54}_{-2.43}$ & -6.38$^{+2.53}_{-2.35}$ & -6.81$^{+2.06}_{-2.04}$ & -5.93$^{+2.48}_{-2.54}$ \\
    H$_{2}$S & [dex]         & -3.37$^{+2.57}_{-4.21}$ & -2.68$^{+1.60}_{-4.68}$ & -4.01$^{+2.97}_{-3.67}$ & -4.27$^{+3.28}_{-3.58}$ & -5.74$^{+2.80}_{-2.69}$ & -3.44$^{+0.77}_{-3.85}$ & -3.96$^{+3.04}_{-3.80}$ \\
    FeH & [dex]              & -5.81$^{+2.72}_{-2.64}$ & -6.06$^{+2.63}_{-2.44}$ & -5.88$^{+2.72}_{-2.48}$ & -5.41$^{+2.75}_{-2.88}$ & -6.84$^{+2.16}_{-2.07}$ & -6.72$^{+2.07}_{-2.10}$ & -5.07$^{+3.03}_{-3.08}$ \\
    TiO & [dex]              & -5.87$^{+2.96}_{-2.64}$ & -5.73$^{+2.91}_{-2.71}$ & -5.71$^{+2.75}_{-2.71}$ & -5.39$^{+2.79}_{-2.80}$ & -6.05$^{+2.38}_{-2.54}$ & -6.59$^{+2.28}_{-2.20}$ & -5.40$^{+2.78}_{-2.89}$ \\
    K & [dex]                & -3.90$^{+2.29}_{-3.92}$ & -4.54$^{+3.14}_{-3.55}$ & -3.11$^{+1.55}_{-4.02}$ & -3.61$^{+2.29}_{-3.78}$ & -5.88$^{+2.49}_{-2.62}$ & -6.03$^{+2.25}_{-2.63}$ & -4.85$^{+3.37}_{-3.20}$ \\
    VO & [dex]               & -5.41$^{+2.96}_{-2.98}$ & -5.34$^{+3.02}_{-3.03}$ & -5.21$^{+2.95}_{-2.97}$ & -5.45$^{+2.86}_{-2.91}$ & -5.05$^{+3.11}_{-3.15}$ & -5.97$^{+2.62}_{-2.58}$ & -5.45$^{+2.91}_{-2.96}$ \\
    C/O & [dex]              & 0.30$^{+0.14}_{-0.11}$ & 0.32$^{+0.15}_{-0.12}$ & 0.25$^{+0.12}_{-0.09}$ & 0.19$^{+0.11}_{-0.08}$ & 1.29$^{+0.41}_{-0.22}$ & 0.35$^{+0.12}_{-0.12}$ & 0.35$^{+0.17}_{-0.14}$ \\
    Fe/H & [dex]             & 1.54$^{+0.57}_{-0.50}$ & 1.05$^{+0.52}_{-0.35}$ & 1.79$^{+0.57}_{-0.57}$ & 1.80$^{+0.54}_{-0.48}$ & 0.97$^{+0.48}_{-0.44}$ & -0.76$^{+0.50}_{-0.44}$ & 2.17$^{+0.48}_{-0.43}$ \\
	\bottomrule
    \end{tabular}
    \label{tab:retrieval_results_CC}
\end{table*}

\begin{table*}[htb!]
    \centering
    \caption{List of posterior median and 68th percentile credible interval for the retrievals based only on LRSP data.}
    \renewcommand*{\arraystretch}{1.7}
    \begin{tabular}{l l|c c c c c c c}
    \toprule
    Parameter & & 1) Baseline & 2) Low T & 3) High T & 4) Low C/O & 5) High C/O & 6) Low Fe/H & 7) High Fe/H \\
    \midrule
    T$_{\mathrm{equ}}$ & [K] & 1744$^{+9}_{-9}$ & 1400$^{+6}_{-6}$ & 2007$^{+15}_{-14}$ & 1738$^{+9}_{-9}$ & 1796$^{+26}_{-25}$ & 1742$^{+19}_{-20}$ & 1739$^{+8}_{-8}$ \\
    $\log(g/g_{0})$ & [dex]  & 4.39$^{+0.06}_{-0.06}$ & 4.31$^{+0.07}_{-0.08}$ & 4.36$^{+0.15}_{-0.19}$ & 4.41$^{+0.10}_{-0.15}$ & 4.60$^{+0.05}_{-0.04}$ & 4.30$^{+0.13}_{-0.15}$ & 4.52$^{+0.26}_{-0.28}$ \\
    $R$ & [R$_{\mathrm{J}}$] & 1.36$^{+0.02}_{-0.02}$ & 1.37$^{+0.01}_{-0.01}$ & 1.35$^{+0.02}_{-0.02}$ & 1.37$^{+0.01}_{-0.01}$ & 1.33$^{+0.04}_{-0.04}$ & 1.35$^{+0.02}_{-0.02}$ & 1.36$^{+0.01}_{-0.01}$ \\
    H$_{2}$O & [dex]         & -1.83$^{+0.03}_{-0.03}$ & -1.80$^{+0.03}_{-0.03}$ & -1.74$^{+0.05}_{-0.05}$ & -1.52$^{+0.03}_{-0.03}$ & -4.05$^{+0.06}_{-0.07}$ & -3.39$^{+0.07}_{-0.08}$ & -0.91$^{+0.06}_{-0.06}$ \\
    CO & [dex]               & -1.67$^{+0.11}_{-0.11}$ & -1.70$^{+0.08}_{-0.08}$ & -1.54$^{+0.34}_{-0.36}$ & -1.68$^{+0.19}_{-0.19}$ & -0.99$^{+0.17}_{-0.20}$ & -3.21$^{+0.13}_{-0.14}$ & -1.24$^{+0.55}_{-0.96}$ \\
    CH$_{4}$ & [dex]         & -7.32$^{+1.78}_{-1.81}$ & -4.16$^{+0.05}_{-0.06}$ & -7.41$^{+1.78}_{-1.72}$ & -6.39$^{+1.98}_{-2.47}$ & -3.05$^{+0.04}_{-0.04}$ & -7.71$^{+1.61}_{-1.56}$ & -6.98$^{+2.14}_{-1.98}$ \\
    CO$_{2}$ & [dex]         & -6.57$^{+2.36}_{-2.25}$ & -6.34$^{+2.44}_{-2.51}$ & -6.54$^{+2.45}_{-2.29}$ & -6.22$^{+2.50}_{-2.49}$ & -7.41$^{+1.92}_{-1.77}$ & -7.31$^{+1.92}_{-1.80}$ & -6.24$^{+2.63}_{-2.42}$ \\
    H$_{2}$S & [dex]         & -5.92$^{+2.70}_{-2.71}$ & -4.60$^{+2.28}_{-3.73}$ & -5.34$^{+3.20}_{-3.13}$ & -5.03$^{+3.14}_{-3.29}$ & -6.81$^{+2.21}_{-2.13}$ & -6.70$^{+2.22}_{-2.25}$ & -5.30$^{+2.97}_{-3.05}$ \\
    FeH & [dex]              & -7.95$^{+0.97}_{-1.40}$ & -9.09$^{+0.62}_{-0.60}$ & -6.66$^{+1.09}_{-2.16}$ & -7.90$^{+1.00}_{-1.40}$ & -6.36$^{+0.16}_{-0.17}$ & -6.85$^{+0.14}_{-0.15}$ & -8.04$^{+0.99}_{-1.23}$ \\
    TiO & [dex]              & -7.82$^{+1.53}_{-1.47}$ & -7.41$^{+1.30}_{-1.70}$ & -4.65$^{+0.43}_{-2.25}$ & -7.67$^{+1.41}_{-1.52}$ & -7.72$^{+1.40}_{-1.55}$ & -7.34$^{+0.99}_{-1.80}$ & -7.20$^{+1.81}_{-1.84}$ \\
    K & [dex]                & -4.06$^{+0.18}_{-0.38}$ & -7.02$^{+1.51}_{-2.04}$ & -3.48$^{+0.59}_{-2.47}$ & -4.17$^{+0.28}_{-1.24}$ & -4.58$^{+0.15}_{-0.15}$ & -5.61$^{+0.38}_{-1.10}$ & -6.34$^{+1.97}_{-2.40}$ \\
    VO & [dex]               & -8.04$^{+0.77}_{-1.25}$ & -9.08$^{+0.67}_{-0.61}$ & -7.56$^{+1.15}_{-1.59}$ & -8.16$^{+0.83}_{-1.18}$ & -8.88$^{+0.78}_{-0.76}$ & -8.98$^{+0.70}_{-0.66}$ & -8.44$^{+0.95}_{-1.00}$ \\
    C/O & [dex]              & 0.48$^{+0.06}_{-0.06}$ & 0.45$^{+0.04}_{-0.04}$ & 0.50$^{+0.17}_{-0.18}$ & 0.31$^{+0.09}_{-0.08}$ & 1.01$^{+0.01}_{-0.00}$ & 0.50$^{+0.06}_{-0.06}$ & 0.23$^{+0.26}_{-0.19}$ \\
    Fe/H & [dex]             & 0.56$^{+0.09}_{-0.08}$ & 0.56$^{+0.08}_{-0.07}$ & 0.71$^{+0.32}_{-0.23}$ & 0.72$^{+0.16}_{-0.11}$ & 1.06$^{+0.20}_{-0.21}$ & -1.00$^{+0.11}_{-0.11}$ & 1.32$^{+0.41}_{-0.24}$ \\
    \bottomrule
    \end{tabular}
    \label{tab:retrieval_results_RES}
\end{table*}

\begin{table*}[htb!]
    \centering
    \caption{List of posterior median and 68th percentile credible interval for the retrievals based on both MRCCS and LRSP data.}
    \renewcommand*{\arraystretch}{1.7}

    \begin{tabular}{l l|c c c c c c c}
    \toprule
    Parameter & & 1) Baseline & 2) Low T & 3) High T & 4) Low C/O & 5) High C/O & 6) Low Fe/H & 7) High Fe/H \\
    \midrule
    T$_{\mathrm{equ}}$ & [K] & 1743$^{+9}_{-9}$ & 1403$^{+6}_{-6}$ & 2010$^{+15}_{-14}$ & 1739$^{+9}_{-9}$ & 1781$^{+23}_{-23}$ & 1753$^{+19}_{-18}$ & 1736$^{+7}_{-8}$ \\
    $\log(g/g_{0})$ & [dex]  & 4.40$^{+0.05}_{-0.05}$ & 4.33$^{+0.07}_{-0.08}$ & 4.42$^{+0.11}_{-0.13}$ & 4.46$^{+0.08}_{-0.09}$ & 4.58$^{+0.03}_{-0.03}$ & 4.32$^{+0.11}_{-0.13}$ & 4.54$^{+0.15}_{-0.16}$ \\
    $R$ & [R$_{\mathrm{J}}$] & 1.36$^{+0.02}_{-0.02}$ & 1.36$^{+0.01}_{-0.01}$ & 1.35$^{+0.02}_{-0.02}$ & 1.37$^{+0.01}_{-0.01}$ & 1.34$^{+0.03}_{-0.03}$ & 1.34$^{+0.02}_{-0.02}$ & 1.37$^{+0.01}_{-0.01}$ \\
    H$_{2}$O & [dex]         & -1.83$^{+0.03}_{-0.03}$ & -1.80$^{+0.03}_{-0.03}$ & -1.77$^{+0.04}_{-0.04}$ & -1.52$^{+0.03}_{-0.03}$ & -4.10$^{+0.07}_{-0.07}$ & -3.38$^{+0.07}_{-0.07}$ & -0.91$^{+0.04}_{-0.04}$ \\
    CO & [dex]               & -1.67$^{+0.11}_{-0.11}$ & -1.68$^{+0.08}_{-0.08}$ & -1.74$^{+0.25}_{-0.24}$ & -1.73$^{+0.15}_{-0.17}$ & -1.08$^{+0.14}_{-0.16}$ & -3.29$^{+0.12}_{-0.12}$ & -0.97$^{+0.27}_{-0.29}$ \\
    CH$_{4}$ & [dex]         & -7.47$^{+1.79}_{-1.67}$ & -4.14$^{+0.06}_{-0.06}$ & -7.49$^{+1.84}_{-1.66}$ & -6.89$^{+1.94}_{-2.08}$ & -3.09$^{+0.03}_{-0.03}$ & -7.46$^{+1.67}_{-1.71}$ & -7.12$^{+2.01}_{-1.90}$ \\
    CO$_{2}$ & [dex]         & -6.72$^{+2.22}_{-2.25}$ & -6.66$^{+2.33}_{-2.21}$ & -6.55$^{+2.25}_{-2.31}$ & -6.56$^{+2.41}_{-2.22}$ & -7.37$^{+1.82}_{-1.80}$ & -7.41$^{+1.85}_{-1.72}$ & -6.29$^{+2.62}_{-2.46}$ \\
    H$_{2}$S & [dex]         & -5.26$^{+3.03}_{-3.26}$ & -2.28$^{+0.37}_{-2.97}$ & -3.33$^{+2.20}_{-4.38}$ & -1.92$^{+0.94}_{-4.97}$ & -6.61$^{+2.24}_{-2.35}$ & -6.34$^{+2.46}_{-2.48}$ & -2.44$^{+1.78}_{-4.69}$ \\
    FeH & [dex]              & -7.86$^{+0.92}_{-1.41}$ & -9.09$^{+0.64}_{-0.60}$ & -7.00$^{+1.27}_{-1.92}$ & -8.03$^{+1.03}_{-1.30}$ & -6.43$^{+0.16}_{-0.17}$ & -6.80$^{+0.13}_{-0.14}$ & -8.07$^{+0.98}_{-1.20}$ \\
    TiO & [dex]              & -7.80$^{+1.50}_{-1.47}$ & -7.30$^{+1.32}_{-1.84}$ & -4.80$^{+0.54}_{-2.92}$ & -7.74$^{+1.56}_{-1.51}$ & -7.83$^{+1.44}_{-1.46}$ & -7.38$^{+1.06}_{-1.79}$ & -7.25$^{+1.73}_{-1.82}$ \\
    K & [dex]                & -4.08$^{+0.18}_{-0.41}$ & -6.71$^{+1.30}_{-2.19}$ & -3.20$^{+0.37}_{-0.77}$ & -4.17$^{+0.22}_{-0.82}$ & -4.61$^{+0.14}_{-0.15}$ & -5.51$^{+0.35}_{-0.85}$ & -6.55$^{+2.06}_{-2.34}$ \\
    VO & [dex]               & -7.97$^{+0.72}_{-1.35}$ & -9.05$^{+0.67}_{-0.64}$ & -7.88$^{+1.28}_{-1.37}$ & -8.28$^{+0.92}_{-1.15}$ & -8.90$^{+0.80}_{-0.76}$ & -9.01$^{+0.72}_{-0.67}$ & -8.42$^{+0.95}_{-1.02}$ \\
    C/O & [dex]              & 0.48$^{+0.05}_{-0.05}$ & 0.46$^{+0.04}_{-0.04}$ & 0.41$^{+0.13}_{-0.11}$ & 0.29$^{+0.07}_{-0.07}$ & 1.02$^{+0.01}_{-0.00}$ & 0.45$^{+0.05}_{-0.05}$ & 0.36$^{+0.14}_{-0.13}$ \\
    Fe/H & [dex]             & 0.58$^{+0.11}_{-0.08}$ & 0.63$^{+0.09}_{-0.08}$ & 0.65$^{+0.45}_{-0.19}$ & 0.81$^{+0.42}_{-0.17}$ & 0.96$^{+0.16}_{-0.17}$ & -1.03$^{+0.11}_{-0.10}$ & 1.57$^{+0.40}_{-0.24}$ \\
    \bottomrule
    \end{tabular}
    \label{tab:retrieval_results_CROCO}
\end{table*}

\begin{table*}[htb!]
    \centering
    \renewcommand*{\arraystretch}{1.7}
    \caption{List of posterior median and 68th percentile credible interval for our retrieval excluding the synthetic VLTI/GRAVITY data (cf. Sect.~\ref{sec:applicability}).}
    \begin{tabular}{l l|c c c}
    \toprule
    Parameter & & MRCCS & LRSP & MRCCS+LRSP \\
    \midrule
    T$_{\mathrm{equ}}$ & [K] & 1900$^{+126}_{-115}$  & 1757$^{+17}_{-17}$ & 1748$^{+15}_{-15}$  \\
    $\log(g/g_{0})$ & [dex]  & 5.55$^{+0.62}_{-0.58}$  & 4.47$^{+0.09}_{-0.13}$ & 4.43$^{+0.09}_{-0.09}$  \\
    $R$ & [R$_{\mathrm{J}}$] & 2.84$^{+2.10}_{-0.87}$  & 1.34$^{+0.02}_{-0.02}$ & 1.35$^{+0.02}_{-0.02}$  \\
    H$_{2}$O & [dex]         & -1.07$^{+0.49}_{-0.38}$  & -1.83$^{+0.10}_{-0.09}$ & -1.87$^{+0.07}_{-0.07}$  \\
    CO & [dex]               & -1.21$^{+0.45}_{-0.42}$  & -4.99$^{+2.93}_{-3.19}$ & -1.85$^{+0.32}_{-0.31}$  \\
    CH$_{4}$ & [dex]         & -5.74$^{+2.44}_{-2.83}$  & -7.53$^{+1.93}_{-1.56}$ & -7.37$^{+1.84}_{-1.76}$  \\
    CO$_{2}$ & [dex]         & -6.17$^{+2.48}_{-2.33}$  & -6.70$^{+2.10}_{-2.13}$ & -6.79$^{+2.15}_{-2.11}$  \\
    H$_{2}$S & [dex]         & -3.37$^{+2.57}_{-4.21}$  & -5.58$^{+2.91}_{-2.86}$ & -2.46$^{+1.09}_{-4.72}$  \\
    FeH & [dex]              & -5.81$^{+2.72}_{-2.64}$  & -8.10$^{+1.03}_{-1.22}$ & -8.14$^{+1.06}_{-1.23}$  \\
    TiO & [dex]              & -5.87$^{+2.96}_{-2.64}$  & -7.80$^{+1.45}_{-1.42}$ & -7.81$^{+1.46}_{-1.43}$  \\
    K & [dex]                & -3.90$^{+2.29}_{-3.92}$  & -4.03$^{+0.23}_{-0.28}$ & -4.05$^{+0.21}_{-0.28}$  \\
    VO & [dex]               & -5.41$^{+2.96}_{-2.98}$  & -8.13$^{+0.81}_{-1.17}$ & -8.19$^{+0.84}_{-1.15}$  \\
    C/O & [dex]              & 0.30$^{+0.14}_{-0.11}$  & 0.00$^{+0.26}_{-0.00}$ & 0.41$^{+0.17}_{-0.16}$  \\
    Fe/H & [dex]             & 1.54$^{+0.57}_{-0.50}$  & 0.20$^{+0.37}_{-0.12}$ & 0.61$^{+0.30}_{-0.23}$  \\
    \bottomrule
    \end{tabular}
    \label{tab:retrieval_results_no_GRAVITY}
\end{table*}

\end{appendix}
\end{document}